\documentclass[useAMS,usenatbib,usegraphicx]{mn2e}
\usepackage{amssymb,multirow}
\usepackage{times}
\newcommand{\Ergs}{erg\,s$^{-1}$cm$^{-2}$\AA$^{-1}$}
\newcommand{\source} {SWIFT~J1753.5--0127}

\newcommand{\msun}{{\rm M}_{\sun}}
\newcommand{\Msun}{M$_\odot$}

\newcommand{\kms}{km\,s$^{-1}$}
\newcommand{\myemail}{vitaly@neustroev.net}
\newcommand{\IRAF}  {{\sc iraf}}

\def\t{$\times$}
\def\l{\ensuremath{\lambda}}
\def\cd{cycle d$^{-1}$}
\newcommand{\tim}[1]{\ensuremath{\times 10^{#1}}}

\newcommand{\Halpha} {H$\alpha$}
\newcommand{\HeI}  {He\,{\sc i}}
\newcommand{\HeII} {He\,{\sc ii}}
\newcommand{\AlII}  {Al\,{\sc ii}}
\newcommand{\NII} {N\,{\sc ii}}
\newcommand{\NIV}  {N\,{\sc iv}}
\newcommand{\SiII} {Si\,{\sc ii}}
\newcommand{\SiIII} {Si\,{\sc iii}}
\newcommand{\SiIV} {Si\,{\sc iv}}

\newcommand{\CI} {C\,{\sc i}}
\newcommand{\CII} {C\,{\sc ii}}
\newcommand{\CIV} {C\,{\sc iv}}

\newcommand{\FeII} {Fe\,{\sc ii}}

\newcommand{\pasp}{PASP}
\newcommand{\apjs}{ApJS}
\newcommand{\apj}{ApJ}
\newcommand{\apjl}{ApJL}

\newcommand{\aap}{A\&A}
\newcommand{\araa}{ARA\&A}
\newcommand{\aj}{AJ}
\newcommand{\mnras}{MNRAS}

\newcommand{\actaa}{AcA}
\newcommand{\memsai}{MmSAI}

\title[System parameters of \source]
{Spectroscopic evidence for a low-mass black hole in \source}
\author[V.~V.~Neustroev et al.]{Vitaly~V.~Neustroev$^{1}$\thanks{\myemail},
Alexandra~Veledina$^{1,2}$, Juri~Poutanen$^{2,1}$, Sergey~V.~Zharikov$^3$,
\newauthor
Sergey~S.~Tsygankov$^{4,1,2}$, George~Sjoberg$^{5,6}$, Jari~J.~E.~Kajava$^{7,8,1}$\\
$^{1}$Astronomy Division, Department of Physics, PO Box 3000, FIN-90014 University
of Oulu, Finland\\
$^{2}$Tuorla Observatory, University of Turku, V\"ais\"al\"antie 20, FIN-21500 Piikki\"o, Finland\\
$^{3}$Instituto de Astronom{\'i}a, Universidad Nacional Aut{\'o}noma de M{\'e}xico, Apdo. Postal 877,
 Ensenada, 22800 Baja California, M{\'e}xico\\
$^{4}$Finnish Centre for Astronomy with ESO (FINCA), University of Turku,
V\"{a}is\"{a}l\"{a}ntie 20, FI-21500 Piikki\"{o}, Finland\\
$^{5}$The George-Elma Observatory, New Mexico Skies, 9 Contentment Crest, \#182, Mayhill, NM 88339, USA\\
$^{6}$American Association of Variable Star Observers, 49 Bay State Road, Cambridge,
MA 02138, USA\\
$^{7}$European Space Astronomy Centre (ESA/ESAC), Science Operations
Department, 28691 Villanueva de la Ca\~{n}ada, Madrid, Spain\\
$^{8}$Nordic Optical Telescope, Apartado 474, 38700 Santa Cruz de La Palma, Spain
}

\begin{document}

\date{Accepted 2014 September 12. Received 2014 September 10; in original form 2014 May 23}

\pagerange{\pageref{firstpage}--\pageref{lastpage}} \pubyear{2014}

\maketitle

\label{firstpage}

\begin{abstract}
The black hole (BH) candidate \source\ has remained active since the onset of its 2005
outburst. Emission lines in the optical spectrum were observed at the very beginning of the
outburst, but since then the spectrum has been featureless making a precise BH mass estimation
impossible. Here we present results from our optical and ultraviolet (UV) observations of \source\
taken in 2012--2013. Our new observations show extremely broad, double-peaked emission lines in the optical
and UV spectra. The optical data also show narrow absorption and emission features with nearly
synchronous and significant Doppler motions. A radial velocity study of these lines which we
associate with the secondary star, yields a semi-amplitude of $K_2$=382 \kms.
A time-series analysis of the spectral and photometric data revealed
a possible orbital periodicity of 2.85 h, significantly shorter than the reported 3.2 h periodic
signal by \citeauthor{Zurita08} The observed variability properties argue against a low orbital
inclination angle and we present several observational arguments in favour of the BH interpretation.
However, the measured radial velocity semi-amplitude of the donor star and the short orbital period
imply that \source\ has one of the lowest measured mass function for a BH in a low-mass X-ray binary.
We show that the compact object mass in excess of 5 \Msun\ is highly improbable. Thus, \source\
is a BH binary that has one of the shortest orbital period and hosts probably one of the smallest
stellar mass BH found to date.
\end{abstract}

\begin{keywords}
accretion, accretion discs -- binaries: close -- stars: individual: \\\source\ -- X-rays: binaries -- X-rays: stars
\end{keywords}

\section{Introduction}

X-ray transients are a subset of the low-mass X-ray binaries (LMXBs) which spend most of their
lives in a quiescent state, with typical X-ray luminosities below $10^{32}$\,erg\,s$^{-1}$.
Occasionally they exhibit bright X-ray and optical outbursts, which occur irregularly with
intervals from a few years to decades or even longer. During the outbursts, the X-ray luminosity
increases by a factor of up to $10^6-10^7$ in a few days and then decays back to quiescence in a few
months.

\source\ is an X-ray transient discovered by the \emph{Swift} Burst Alert Telescope (BAT) on 2005 May 30
as a bright variable X-ray source \citep{Palmer05}. Although the mass of the primary has not been
dynamically measured, the system displays a number of characteristics that suggests the binary hosts a
black hole (BH). First, the hard X-ray spectrum at the outburst  peak had a  maximum
(in $\nu F_{\nu}$) at $\sim$150~keV \citep{Cadolle07}, while the corresponding peaks in neutron star (NS)
binaries do not exceed 50~keV \citep{Barret00,PG03,GP05,Lin07,Lin10,IP09}.
Secondly, the X-ray spectrum of \source\ has significantly hardened during the decline phase,
with the photon spectral index reaching values as low as $\Gamma=1.65$ \citep{Chiang10}, which
is typical for hard state BHs, but  is much smaller than that in binaries hosting a NS \citep{ZPM98,ZG04,Gilfanov10}.
Thirdly, the X-ray power density spectrum of \source\
reveals a strong power suppression at frequencies above $\sim$10~Hz  \citep{Durant09,Soleri13},
unlike in NS binaries, which show significant power above $\sim$500~Hz \citep{SR00}.
Moreover, temporal analysis of the \textit{RXTE} data  \citep{Morgan05}
revealed the presence of the low-frequency quasi-periodic oscillations (QPOs) with a shape typically seen in the BH candidates
\citep[type C QPO, see ][]{Belloni11}.
Neither X-ray bursts nor pulsations were detected since the initial outburst further hinting towards
the BH nature of the central source in the binary.

It is intriguing that 9 yr after the beginning of the outburst, \source\ has not yet returned to
the quiescent state. This unusual behaviour has triggered an interest in the binary.
Among other properties, a challenging task was to measure system parameters. Neither the mass $M_1$
of the primary nor even the orbital period $P_{\rm orb}$ is reliably measured at the moment.
The optical photometry conducted by \citet{Zurita08} and \citet{Durant09} revealed a light curve
with a complex non-sinusoidal morphology. The detected $\sim$3.24~h modulations were attributed
to the superhump period, which is known to be slightly longer than the orbital period $P_{\rm orb}$
\citep{Patterson05}. This makes \source\ one of the shortest orbital period BH systems.

The most direct method of estimating the system parameters of the binary system involves optical spectroscopy,
which can give the orbital radial velocity curve of the donor star and of the accretion disc around the primary
compact object. A precise determination of the radial velocity amplitude of the secondary in a LMXB is of the
highest importance as it allows us to set an absolute lower limit for the mass of the compact object.
These measurements are usually done through the analysis of absorption spectra of the secondary.
The most appropriate time to study the nature and dynamical properties of the secondary star in
X-ray transients is quiescence, during which light from the star contributes significantly to the visible spectrum.
Unfortunately, for the case of \source\ it will be a complicated task owing to the faintness of the
source during a quiescent state. It is not visible in archival images and can be as faint as
$V$$\sim$21 mag \citep{Cadolle07}.

However, some dynamical information about the secondary star can potentially be derived even during outburst.
For example, \citet{SC02} detected a large number of very narrow emission line features from the irradiated
secondary star in the persistent X-ray binary Sco X-1. Furthermore, the discovery of sharp emission
components of the Bowen blend in  Sco X-1 and the X-ray transient GX~339--4 caught in an outburst state
allowed the determination of the primary masses \citep{SC02,CSH03,HSC03,munoz08}.

No time-resolved spectroscopic studies for \source\ were reported to date, even though a few spectra
were presented in the past. The observations near the outburst peak revealed the presence of
the double-peaked \Halpha\ and \HeII~\l4686 lines \citep{Torres05b}, which seemed to
disappear a month later  \citep{Cadolle07}. After that the object showed a rather featureless
spectrum \citep{Durant09}. Our photometric monitoring of \source\ revealed that in 2012 the binary
has weakened by at least $\sim$0.3--0.4 mag in comparison with the previous spectroscopic observations.
This gave us grounds to suspect that the lines might have become more apparent. This motivated us to
perform time-resolved spectroscopy of \source\ in order to attempt detection of these lines and to
estimate the system parameters. Here we present a study of \source\ based on our optical and ultraviolet
(UV) spectroscopic observations, supported by the optical photometric monitoring.

%-----------------------------Table Start--------------------------------
\begin{table*}
\label{ObsTab}
\begin{center}
%\begin{flushleft}
\caption{Log of optical time-resolved observations of \source.}
\begin{tabular}{cclcrrc}
\hline\hline
 Date        & HJD Start  & Telescope /   & Filter /        & Exp.Time & Number   & Duration\\
             &  2450000+  & Instrument    & $\lambda$~range (\AA)  &  (s)   & of exps. &  (h)\\
\hline
2013-Aug-06  &  6510.808  & 2.1~m / B\&Ch & 3600--7000      &  600     & 12       & 1.97   \\
2013-Aug-07  &  6511.688  & 1.5~m / RATIR & $V$, $i$        &  80      & 140      & 3.78   \\
             &  6511.718  & 2.1~m / B\&Ch & 3600--7000      &  900     & 12       & 3.66   \\
2013-Aug-08  &  6512.660  & 2.1~m / B\&Ch & 3600--7000      &  900     & 18       & 5.15   \\
             &  6512.684  & 1.5~m / RATIR & $V$, $i$        &  80      & 140      & 3.62   \\
2013-Aug-09  &  6513.712  & 2.1~m / B\&Ch & 3600--7000      &  900     & 12       & 3.84   \\
\hline
\end{tabular}
\end{center}
\end{table*}
%-----------------------------Table End--------------------------------

\section{Observations and data reduction}

\subsection{\emph{HST} Ultraviolet Observations}
\label{HSTobsSec}

We observed \source\ with the Cosmic Origins Spectrograph (COS) aboard \emph{Hubble Space Telescope}
(\emph{HST}) on 2012 October 8
(PID 12919). The far-UV (FUV) spectra were collected using the low-resolution grating
G140L in the 1280\,\AA\ setting. This grating covers 1260--2400\,\AA\ and 200--1170\,\AA\ on the
A and B segments of the detector, respectively, with  a  spectral resolution of $\sim$0.75\,\AA.
The total exposure time of the observations was 1.36 h acquired over a 2.2 h time period (two orbits).

The data were analysed using {\sc pyraf} routines from {\sc stsdas} package {\sc hstcos} (version 3.16).
The data from each of the four exposures were summed to produce a single spectrum.
Because of very low instrument sensitivity at shorter and longer wavelengths and the relatively low flux
of the object ($f_{\lambda}\lesssim 2\times 10^{-15}$\Ergs), we cut out the B segment of the spectrum at
1080--1170\,\AA\ and the A segment at 1260--2000\,\AA.

We also used another data set of the COS observations of \source\ retrieved from the Multimission
Archive at STScI (MAST) and reduced in a similar way to that above. These data (PID 12039) taken in
both the FUV and near-UV (NUV) channels, were obtained 6 d before our observations
during  five \emph{HST} orbits, on 2012 October 2.
The FUV spectra in this data set were taken with the high-resolution gratings G130M and G160M.
The reduced spectra closely resemble our spectrum but are much noisier. Therefore, we
did not use them for the following analysis.
In the NUV region, the low-resolution grating G230L in the 2950\,\AA\ and 3360\,\AA\ settings was
used, with the exposure time of $\sim$0.4 h in each setting. Four NUV segments were covered
(1690--2090\,\AA, 2120--2515\,\AA, 2790--3180\,\AA\ and 3205--3600\,\AA) providing a
spectral resolution of $\sim$0.8\,\AA.
The G230L data for wavelengths longwards of 3200\,\AA\ can be strongly
contaminated by second-order light and the flux calibration applied by {\sc calcos} at these
wavelengths is unreliable \citep{COSreport}. For this reason we excluded  from our analysis
the segment covering the wavelength range 3205--3600\,\AA.

%-----------------------------Table Start--------------------------------
\begin{table*}
\caption{Log of photometric observations of \source.}
\label{ObsPhotTab}
\begin{tabular}{ccccccccc}
\hline
\hline
  Date        &  HJD start &  Exp. Time     & Duration & \multicolumn{4}{c}{Average Magnitudes} \\
              &  2450000+  &     (s)        &   (h)    &       $B$      &       $V$      &       $R$      &       $I$      \\
\hline
 2012-Sep-21  &  6191.687  &    300         &   1.33   & 17.30$\pm$0.04 & 16.90$\pm$0.03 & 16.63$\pm$0.02 & 16.16$\pm$0.02 \\
 2012-Oct-03  &  6203.630  &    420         &   1.40   &    \dots       & 16.88$\pm$0.03 & 16.62$\pm$0.02 & 16.19$\pm$0.02 \\
 2012-Oct-08  &  6208.632  &    300         &   1.67   &    \dots       &       \dots    & 16.58$\pm$0.01 &      \dots     \\
 2012-Oct-09  &  6209.589  &    300         &   0.67   & 17.18$\pm$0.05 & 16.85$\pm$0.04 & 16.66$\pm$0.03 & 16.15$\pm$0.03 \\
 2013-Apr-12  &  6394.903  &    300         &   1.33   & 17.30$\pm$0.04 & 16.89$\pm$0.03 & 16.63$\pm$0.02 & 16.10$\pm$0.02 \\
 2013-Apr-16  &  6398.926  &    300         &   1.33   & 17.28$\pm$0.04 & 16.91$\pm$0.03 & 16.64$\pm$0.02 & 16.19$\pm$0.02 \\
 2013-Jun-19  &  6462.728  &    300         &   3.33   & 17.24$\pm$0.05 & 16.93$\pm$0.02 & 16.62$\pm$0.03 & 16.19$\pm$0.03 \\
 2013-Aug-09  &  6513.631  &    300         &   3.67   & 17.33$\pm$0.05 & 16.93$\pm$0.02 & 16.63$\pm$0.03 & 16.15$\pm$0.03 \\
 2013-Aug-10  &  6514.627  &    300         &   3.33   & 17.33$\pm$0.05 & 16.91$\pm$0.02 & 16.67$\pm$0.03 & 16.17$\pm$0.03 \\
 2013-Aug-11  &  6515.543  &    300         &   3.33   & 17.31$\pm$0.05 & 16.89$\pm$0.03 & 16.65$\pm$0.03 & 16.12$\pm$0.03 \\
\hline
 Mean         &            &                &          & 17.28$\pm$0.05 & 16.90$\pm$0.02 & 16.63$\pm$0.03 & 16.16$\pm$0.03 \\
\hline
\end{tabular}
\end{table*}

%-----------------------------Table End--------------------------------

%-----------------------------Figure Start------------------------------
\begin{figure*}
\centerline{
  \includegraphics[width=8.5cm]{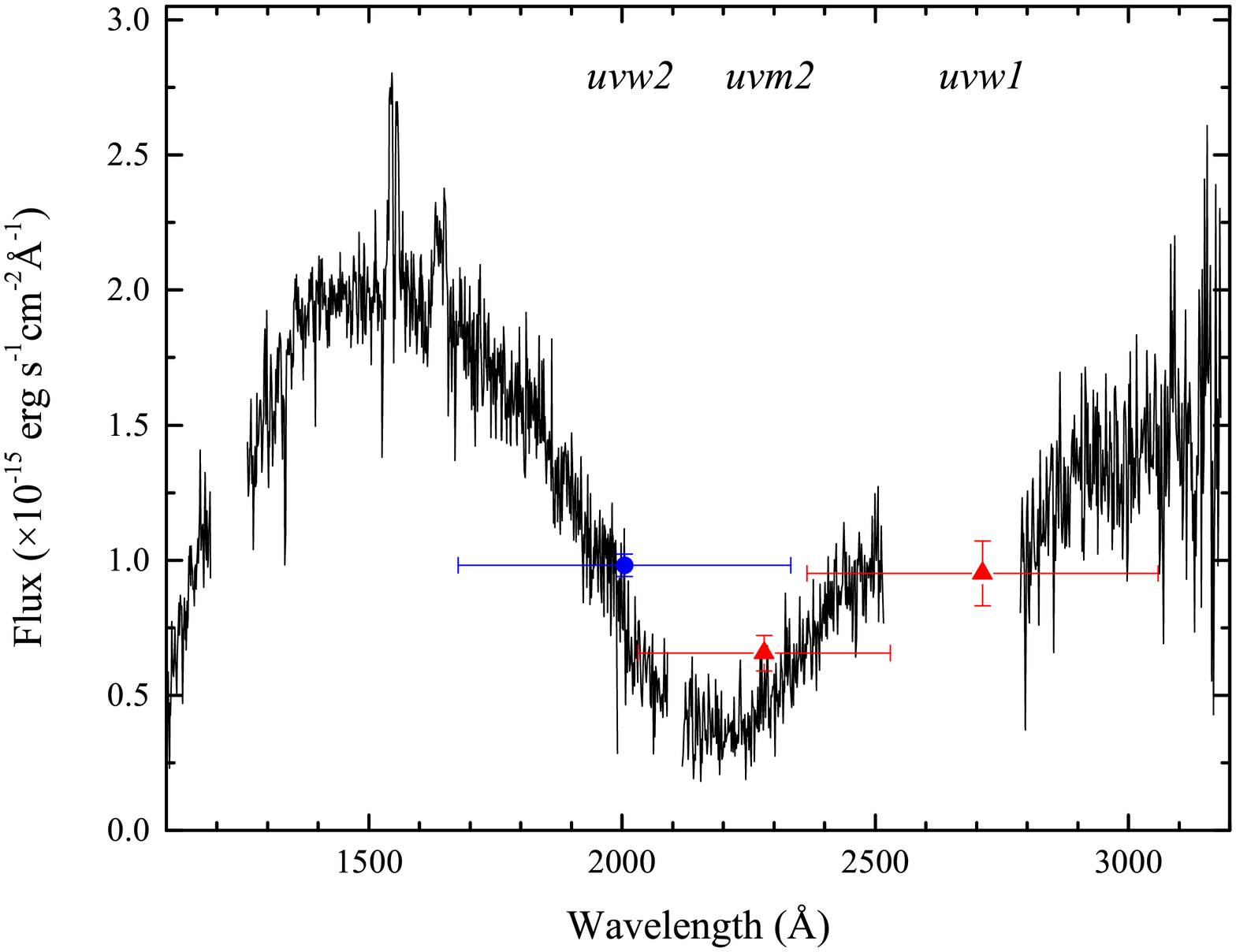}
  \includegraphics[width=8.5cm]{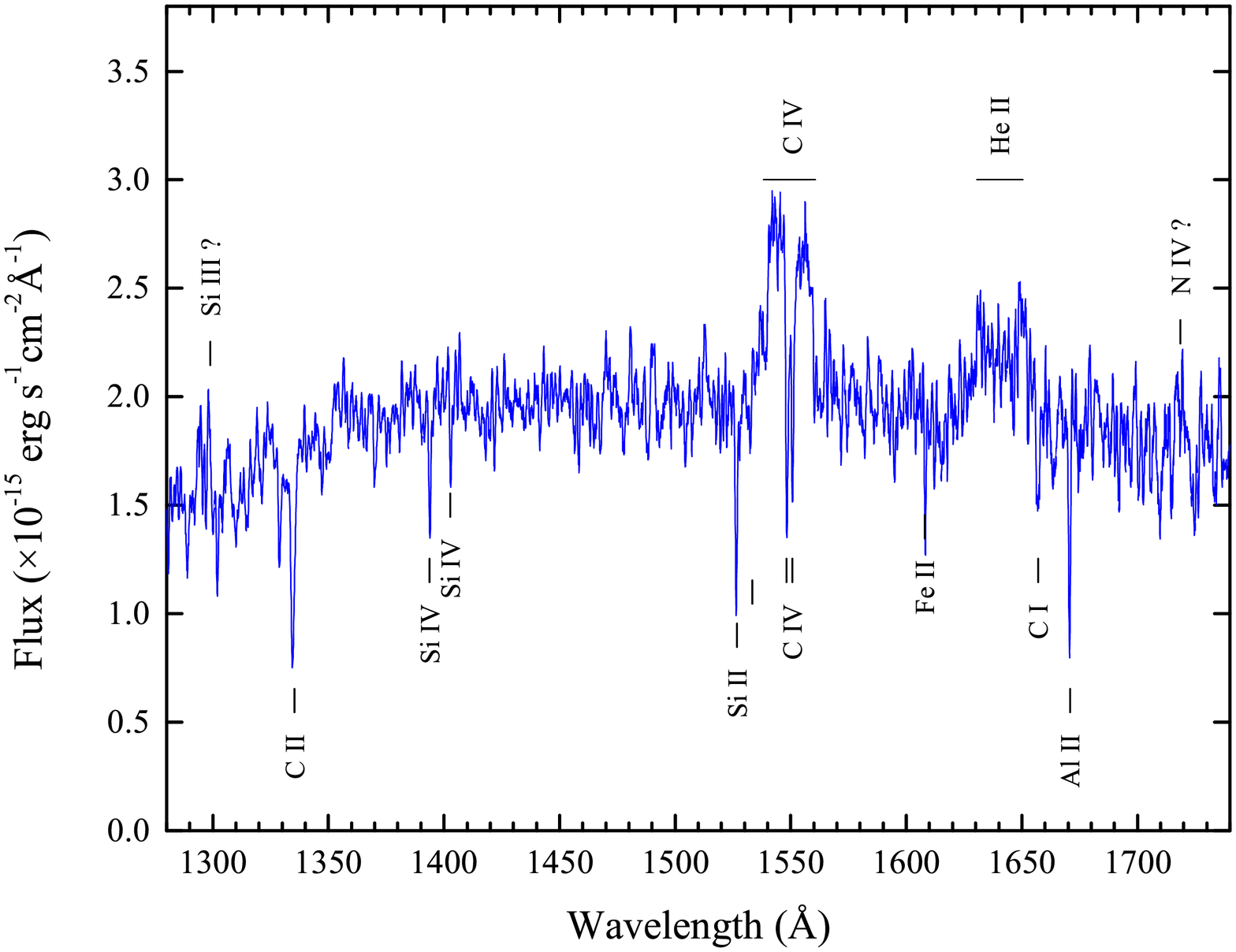}
}
\caption{Left: The UV spectrum of \source\ uncorrected for interstellar reddening. The data have been
         binned to $\sim$2.5~{\AA} spectral resolution. The blue circle and red triangles show the Swift/UVOT measurements.
         Right: A part of the FUV spectrum containing the strongest lines and smoothed with a 20 point
         running boxcar to show the weaker lines.
         }
\label{Fig:UV_spec}
\end{figure*}
%-----------------------------Figure End--------------------------------

\subsection{Optical time-resolved observations}
\label{ObsSec}

  The optical spectra of \source\ were obtained during four consecutive nights of 2013 August 6--9
  at the Observatorio Astronomico Nacional (OAN SPM) in Mexico on the 2.1-m telescope.
  The observations were conducted with the Boller and Chivens spectrograph, equipped with a
  13.5~$\mu$m ($2174\times2048$) Marconi E2V-4240 CCD chip. A total of 54 spectra were obtained
  in the wavelength range of 3600--7000 {\AA}, in the first order of a 400 line mm$^{-1}$ grating with
  corresponding spectral resolution of $\sim$4.5\,\AA\ measured from the night-sky lines.
  During the first night of observations
  600~s individual exposures were taken (12 spectra), while the rest of the spectra were obtained
  using 900~s individual exposures. The total exposure time was 750 min that allowed us to achieve
  a signal-to-noise ratio (SNR) in the averaged spectrum of $\sim$160 at 5200 \AA.

  All the nights of observations were photometric with exception for the first half of the second
  night (August 7) when cirrus clouds could have affected the flux level of the output spectra.
  The seeing ranged from 1 to 2 arcsec. The slit width of 2.5 arcsec was chosen to avoid slit
  losses due to possible imperfect pointing of the telescope.
  We note that because of the employed relatively wide slit, seeing and telescope pointing variations could
  lead to a variable slit illumination profile of the target, which in turn could lead to corresponding
  radial velocity shifts in the dispersed spectra. However, this should not be a significant problem since
  in the following analysis we use phase averaged data which show no significant spurious radial velocity
  shifts.
  In order to apply an accurate flux correction, three
  standard spectrophotometric stars were observed every night. They were selected from Feige\,110,
  HZ\,44, BD+33\,2642 and BD+28\,4211 \citep{Oke}. Comparison spectra of Cu-He-Ne-Ar lamp were used
  for the wavelength calibration. The data reduction was performed using the \IRAF\ environment.

  During the nights of August 7 and 8, the spectroscopic observations were accompanied by photometric
  time-resolved observations on the Harold L. Johnson 1.5~m telescope at the same site. The observations
  were performed simultaneously in the Johnson $V$ and SDSS $i$ bands with the use of a multichannel
  imager {RATIR} \citep{Ratir1,Ratir2}. The exposure times were 80~s. The $V$ magnitudes of the
  object were determined using the calibration stars reported by \citet{Zurita08}. Because of the lack of
  observations of photometric standard stars in the SDSS $i$ filter, we used an arbitrary zero-point
  for the $i$ measurements.
  Table~\ref{ObsTab} provides the journal of the optical time-resolved observations of \source.

%-----------------------------Figure Start------------------------------
\begin{figure*}
\centerline{
  \includegraphics[width=16cm]{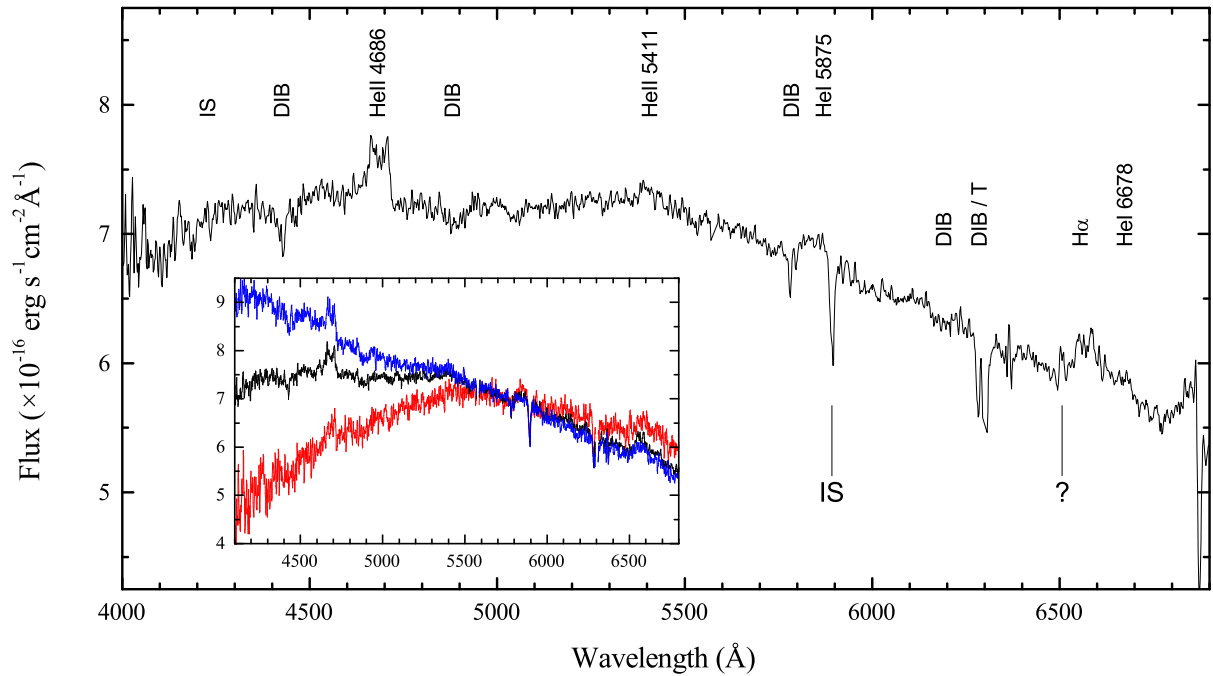}
}
\caption{The averaged optical spectrum of \source. The inset shows the extreme shapes of the variable continuum.}
\label{Fig:opt_spec}
\end{figure*}
%-----------------------------Figure End--------------------------------

%-----------------------------Figure Start------------------------------
\begin{figure*}
\centerline{
  \includegraphics[height=7.5cm]{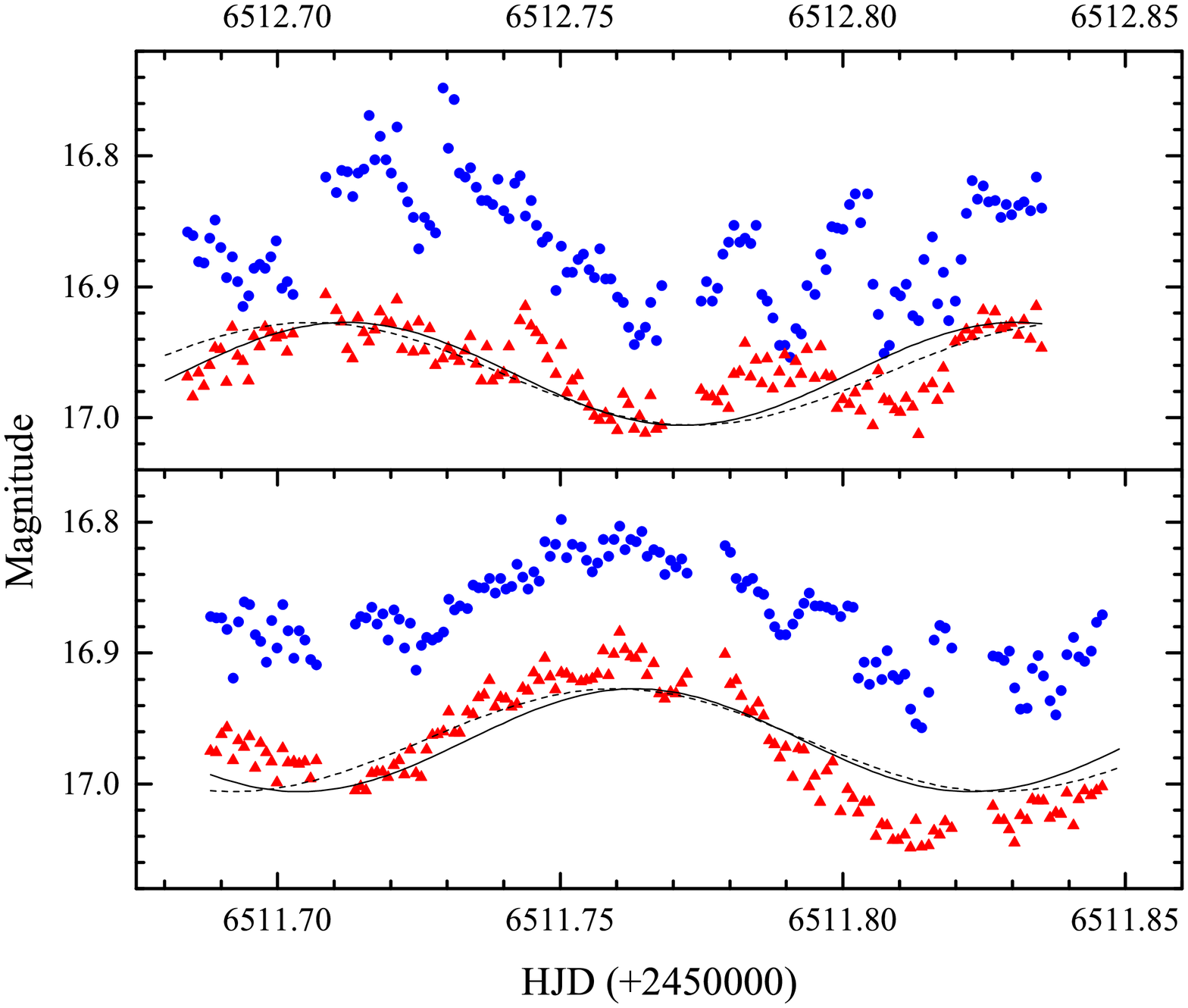}
  \includegraphics[height=7.5cm]{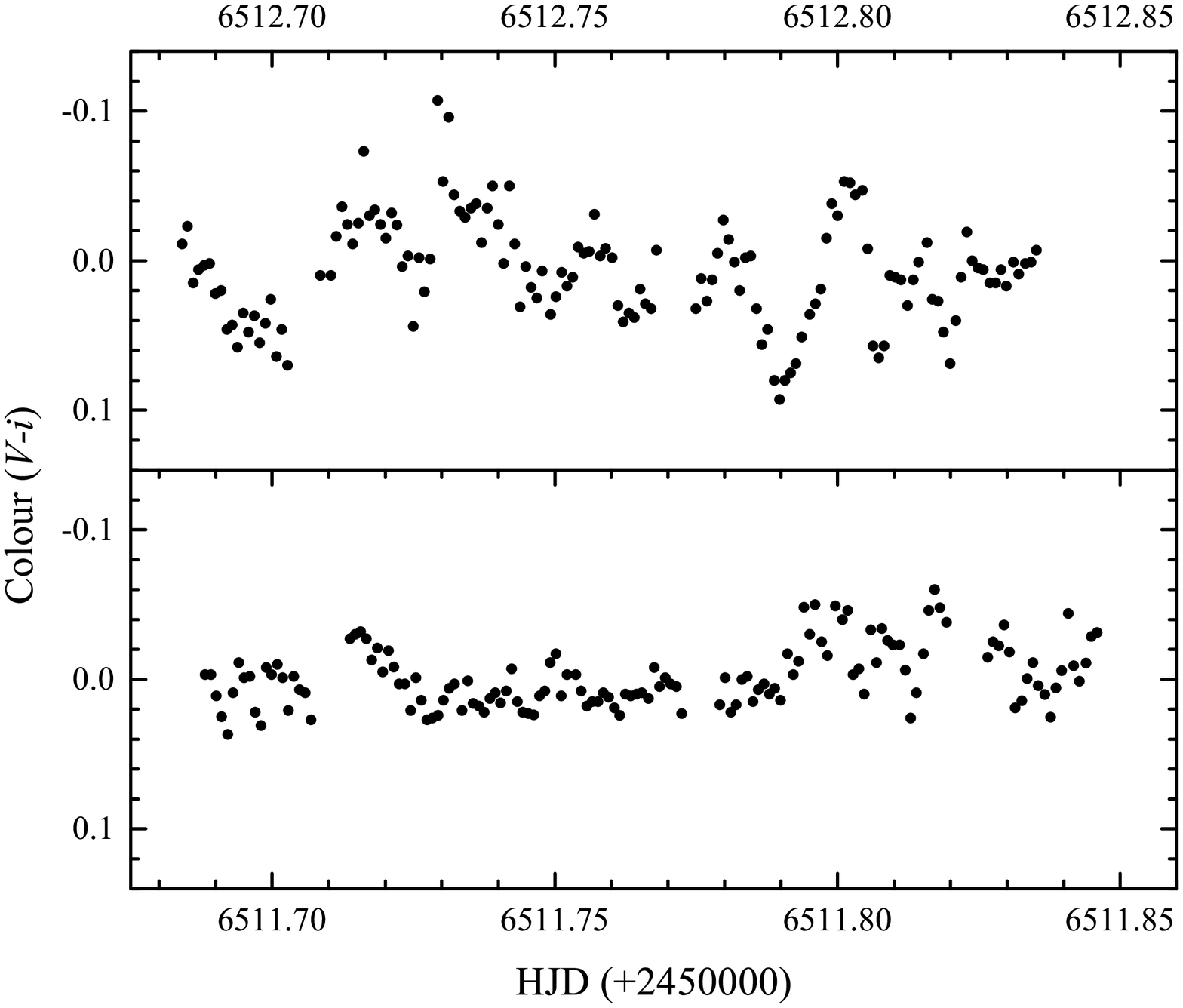}
}
\caption{Left: the $V$ (blue circles) and SDSS $i$-band (red triangles) light curves of \source.
Because of the lack of observations of photometric standard stars in the SDSS $i$ filter, an
arbitrary zero-point was used for the $i$-band measurements. The solid and dashed lines are
sinusoidal fits to the $i$-band data with the period constrained to $\sim$2.85~h
and $\sim$3.24~h, respectively.
Right: the corresponding $V-i$ colour
(with the linear trend subtracted for each night of observations).
}
\label{Fig:Ratir}
\end{figure*}
%-----------------------------Figure End--------------------------------

\subsection{Optical and UV photometric monitoring}

In order to investigate the long-term photometric behaviour and to check for the current state of the
object during our spectroscopic observations, we additionally obtained several sets of
multicolour photometric data. These observations were performed from 2012 September to 2013 August
at New Mexico Skies in Mayhill, New Mexico, with the 0.35-m Celestron C14 robotic telescope
and an SBIG ST-10XME CCD camera with Johnson-Cousins $BV(RI)_C$ Astrodon Photometric filters.
The images were usually taken in a sequence $B$-$V$-$R$-$I$ with exposure
times of 300 s for each filter. In order to establish nightly averaged fluxes, we observed
the object for several hours per night.
The data reduction  was performed using the \IRAF\ environment and the software AIP4Win
v.~2.4.0 \citep{AIP4Win}. In order to improve the confidence of our measurements we
aligned and summed all the nightly images for each filter and then measured the average magnitudes.
We used the secondary standards found in \citet{Zurita08} to establish the zero points.
Table~\ref{ObsPhotTab} provides a journal of the photometric observations with the measured
magnitudes. We note that the $VRI$ magnitudes obtained on 2012 October 3 are very close to
the values of \citet{Froning} obtained a day before.

We also performed a Target of Opportunity (ToO) observation of \source\ in  $uvw2$ filter
with the UV-Optical Telescope
(UVOT)  onboard the {\it Swift} X-ray satellite \citep{Swift} on 2012 September 19.
Furthermore, we used the  {\it Swift}/UVOT data obtained during the entire 2012 year.
These observations were analysed with the \emph{Swift} Release 3.7
software\footnote{http://swift.gsfc.nasa.gov/docs/software/lheasoft/} together
with the most recent version of the Calibration Database.
The UVOT observations were reduced following the procedure described in \citet{UVOT}.

\section{Data analysis and results}

\subsection{UV spectrum}

The  UV spectrum of \source\ is shown in Fig.~\ref{Fig:UV_spec} (left-hand panel).
The spectrum is uncorrected  for the interstellar reddening, which manifests itself in a deep
absorption feature centered at 2175\,\AA. We also show the UVOT measurements in UV filters
(\textit{uvw1}, \textit{uvm2}, \textit{uvw2}). Only the \textit{uvw2}
measurements were obtained near the time of the \emph{HST} observations, the fluxes in other UVOT
filters are the average of the measurements obtained during the year 2012. The UVOT fluxes are
very similar to those derived from the spectroscopy.

The UV spectrum is dominated by broad and double-peaked emission lines of \CIV\ \l1550
and \HeII\ \l1640 with no P Cyg absorption components. These properties allow us to place constraints
on the orbital inclination (see Section~\ref{Sec:Inclination}), and on the accretion disc outer
radii where the UV emission lines originate.
There is also a hint of other weaker emission lines
of \SiIII\ $\lambda1298$ and \NIV\ $\lambda1718$.
The absorption spectrum is rich in features, but all these absorption lines are very narrow and
are not shifted with respect to the rest wavelength being consistent with the
interstellar origin. The strongest lines are \CI\ \l1657, \CII\ \l1335, \CIV\ (\l1548.2, \l1550.7),
\SiII\ (\l1527, \l1533), \SiIV\ (\l1393.8, \l1402.7), \FeII\ \l1608, \AlII\ \l1671.
In the NUV spectrum we were unable to identify any spectral lines.
A part of the FUV smoothed spectrum containing the strongest lines
is shown in Fig.~\ref{Fig:UV_spec} (right-hand panel).

\subsection{Optical spectrum}
\label{Sec:OptSpec}

The averaged optical spectrum of \source\ is shown in Fig.~\ref{Fig:opt_spec}. It exhibits very
broad double-peaked emission lines of \HeII\ \l4686 and \Halpha. There are also hints of
\HeII\ \l5411 and \HeI\ emission lines (5875 \AA, 6678 \AA) and of the Bowen blend. We also
note the presence of a narrow, unidentified emission line at $\sim$6507\,\AA.
This feature seems to be real and not an artefact of the data reduction because it is present
in all our subsets of observations of \source, but is not seen in spectra of any other targets
observed the same nights. To the best of our knowledge, the detection of a similarly weak
emission line in this wavelength region was reported only for the X-ray transient GX~339--4.
\citet{Line6507} identified this line with \NII\ \l6505. However, we doubt this identification
as no other \NII\ lines are observed in either \source\ or GX~339--4.

Several diffuse interstellar bands (DIB), interstellar (IS) and telluric absorption lines (T) are
apparent in the spectrum. Besides them, no other absorption lines which might be identified with
the secondary star are seen in this averaged spectrum. However, the time-resolved spectra reveal
a few narrow absorption and emission features which show significant sinusoidal Doppler
shifts. We discuss these features in Section~\ref{Sec:radialsec}.

The continuum in the averaged spectrum has a broken-line shape with a knee around 5400\,\AA,
somewhat similar to the WHT/ISIS spectrum obtained in 2006 June 17  \citep{Durant09}, but
in contrast to most of other spectra from early observations
which could be reasonably well represented by a straight line  \citep{Cadolle07,Durant09}.
A visual inspection of individual spectra has shown that the continuum shape exhibits
strong variability. The slope of the shorter wavelength segment significantly varies,
while the longer wavelength segment is much more stable (see the inset in
Fig.~\ref{Fig:opt_spec}). We checked if these variations could be caused by wavelength
dependent slit losses and found no correlation with seeing and airmass. Moreover,
the spectra of other variable and standard stars taken
during the same nights with the same telescope/instrument  did not reveal such
variability of the continuum shape. Thus, the detected variability is real
and is not an artefact of the flux calibration.

In order to study the time dependence of this variability, we calculated
the ratio $F_{4300}/F_{6250}$, where $F_{4300}$ and $F_{6250}$ are the fluxes
averaged across the wavelength bands centered at 4300\,\AA\ and 6250\,\AA\ with the widths of
600\,\AA\ and 500\,\AA, respectively. This ratio is analogous to the colour $B-R$.
In Section~\ref{Sec:Period}, we show that $F_{4300}/F_{6250}$ displays a notable variability
with the orbital period. In addition, there are signs in our {RATIR} data of another, shorter
time-scale  variability associated with varying continuum shape, which is stronger in the
$V$-band than in the $i$-band. The corresponding peak-to-peak amplitude of the $V-i$ colour index
variability is as large as $\sim$0.15~mag, which is similar to the amplitude of the orbital
modulation (Fig.~\ref{Fig:Ratir}). Because of the shortness of these {RATIR} observations we are
unable to reach a firm conclusion on the periodicity of the detected variability, but the data
hint towards the period of  $\sim33$ min.
Longer multicolour observations with high enough time resolution preferably in the $B$ and $R$
filters should shade more light on this phenomenon. In conclusion, we note that the nightly
averaged magnitudes are very stable in all the filters except for $B$ (Table~\ref{ObsPhotTab};
see also \citealt{Froning}). The latter shows the relatively large observed scatter in magnitudes.
It can naturally be explained by the optical continuum variability if it is also present on
the longer time scales.

\subsection{Orbital period}
\label{Sec:Period}

\begin{figure}
\centerline{
  \includegraphics[width=8.5cm]{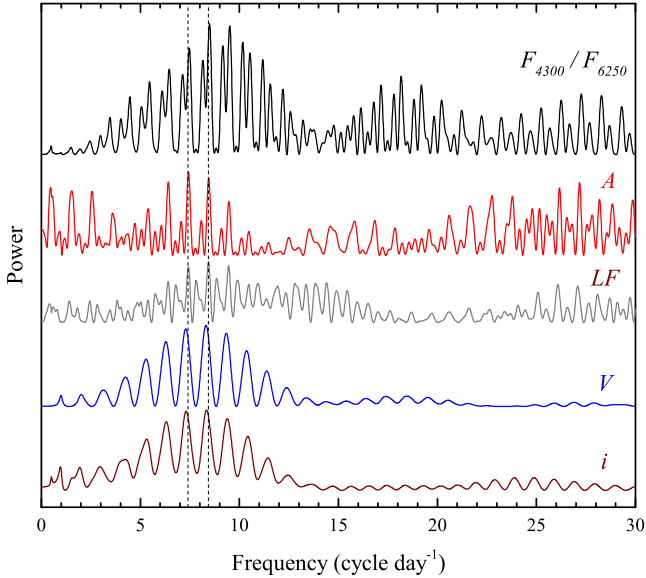}
}
\caption{Power spectra of different quantities calculated from the optical photometric and
         spectroscopic data (see text for explanation). The vertical dashed lines mark
         the adopted orbital frequency of 8.41 \cd\ and the previously proposed 3.24~h period
         (7.41 \cd).
}
\label{Fig:PS}
\end{figure}

\source\ is known to show relatively strong modulations as seen in its optical light curve
\citep{Zurita08, Durant09}, however the precise orbital period of the system is still
unknown. \citet{Zurita08} reported a determination of the modulation period to be
$\sim$3.24~h which they attributed to a superhump period. During our observations,
a similar modulation is also clearly visible in both $V$ and $i$ filters
(Fig.~\ref{Fig:Ratir}).
There also is an apparent colour variability that is in contrast to the observations of
\citet{Durant09}, during which the colours varied very little.
The Lomb-Scargle periodograms for the $V$ and $i$ RATIR photometric data show the
strongest peaks at a frequency of $\sim$8.4 cycle day$^{-1}$ (Fig.~\ref{Fig:PS}),
which is close to the 1-d alias of the period found by \citet{Zurita08}.

Besides the analysis of  photometric variability, we also performed a time series analysis
of our optical spectroscopic data. In addition to the colour index $F_{4300}/F_{6250}$
introduced in Section~\ref{Sec:OptSpec}, we also calculated additional quantities.
The quantity LF (normalized line flux) is closely related to the equivalent width (EW)
of \HeII\ $\lambda$4686\ and defined as the flux integrated over the wavelength range
\l\l4645--4726\,\AA\ divided by the averaged flux in the wavelength ranges
\l\l4512--4612\,\AA\ and \l\l4727--4827\,\AA. We also introduce the asymmetry parameter
$A$ as the ratio of the EWs of the \HeII~\l4686 line on each side of the
rest wavelength. This parameter is sensitive to the variations of both
the radial velocity and the shape of the \HeII\ \l4686 line profile.

The power spectra of $F_{4300}/F_{6250}$, $A$ and $LF$ are also shown in Fig.~\ref{Fig:PS}.
Most of them are dominated by a peak at the same frequency of $\sim$8.4 \cd, with
its averaged value being $8.41\pm0.04$ \cd\ (2.85$\pm0.01$~h), where the error is
expressed as the standard deviation of the mean. The existence of the
variability with the same period in both the photometric and spectroscopic
data indicates that this is the true orbital period.

%-----------------------------Figure Start------------------------------
%-----------------------------Figure End--------------------------------

%-----------------------------Figure Start------------------------------
\begin{figure}
\centerline{
  \includegraphics[width=8.5cm]{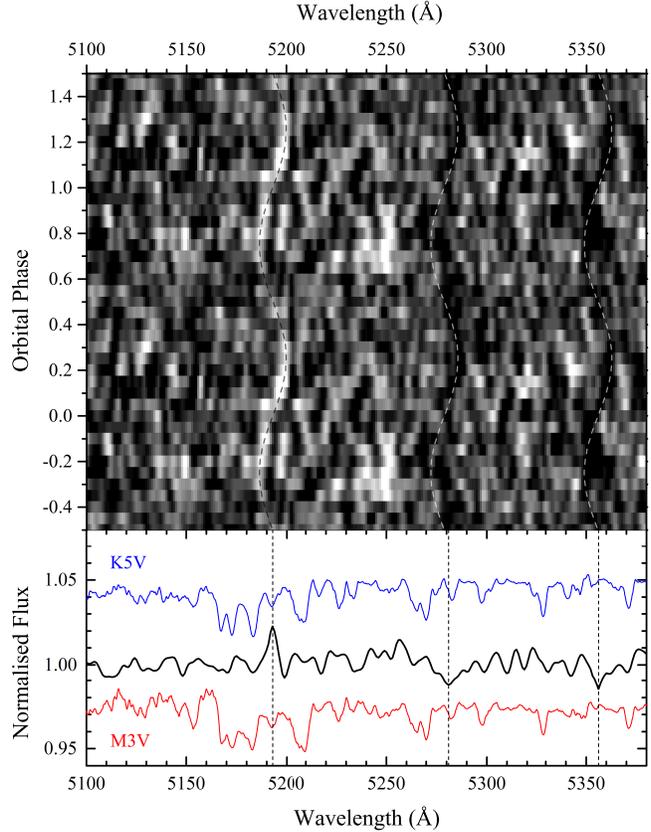}
}
\caption{Top: a part of the trailed spectrum of \source\ showing sinusoidal trails of absorption
         and emission features, marked by the dashed lines. White indicates emission. Two cycles
         are shown for clarity.
         Bottom: the averaged and normalized spectrum of \source, compared with two spectral standards.
         The spectrum of \source\ have been corrected for orbital motion of the secondary star.
         }
\label{Fig:TrailedSpec}
\end{figure}
%-----------------------------Figure End--------------------------------

\subsection{The radial velocity of the secondary star}
\label{Sec:radialsec}

In outburst, the optical flux from the binary system is dominated by emission from the luminous
accretion disc, making the photospheric absorption lines of the donor virtually undetectable.
However, closer inspection of the trailed spectrum of \source\ revealed hints of several narrow
absorption and emission features that showed radial velocity variations with a large amplitude
and in phase with each other. These weak features are more or less clearly seen only in
the wavelength region with the highest SNR ($\sim$5100--5400 \AA).
A few other absorption lines may also be presented in other spectral regions, though their detection is less reliable.
The strongest detected lines are located at $\sim$5193.0\,\AA, 5279.0\,\AA\ and 5356.1\,\AA\ and marked by the dashed
lines in Fig.~\ref{Fig:TrailedSpec} (top panel).
Surprisingly, their identification is unclear. This spectral region of late K- and M-type stars
contains a wealth of absorption lines. Some of them can be matched with the features detected in
the spectrum of \source, but the absence of other expected lines stalls the identification
(Fig.~\ref{Fig:TrailedSpec}, bottom panel).

Nevertheless, the presence in the spectrum of an ensemble of spectral
features with nearly synchronous and significant Doppler motions indicates their common origin
which we tentatively associate with the secondary star.

In order to characterize these velocity variations, we first phased the individual spectra with
the orbital period of 2.85~h and then co-added the spectra into 20 separate phase bins. We placed
orbital phase zero at HJD~245\,6510.8081 -- the time of inferior conjunction of the secondary star
(derived later in this section). The resulted phase-folded spectra have relatively high SNR of
$\sim$50--70 (at 5200 \AA).

The radial velocity measurements of the donor star in LMXBs are usually obtained through
cross-correlation of the absorption lines with stellar templates of similar spectral type. This
method works even if the spectral features of the secondary star are not obvious in the spectrum.
However, the mismatch between spectral lines of \source\ and the standard stars requires to use
another template. Following the iterative approach described in \citet{BF_Eri}, we created the
cross-correlation template spectrum from the observed spectra of the system. This approach
maximizes the similarity between the template and the individual spectra to be cross-correlated.

\begin{figure}
\centerline{
  \includegraphics[width=8.0cm]{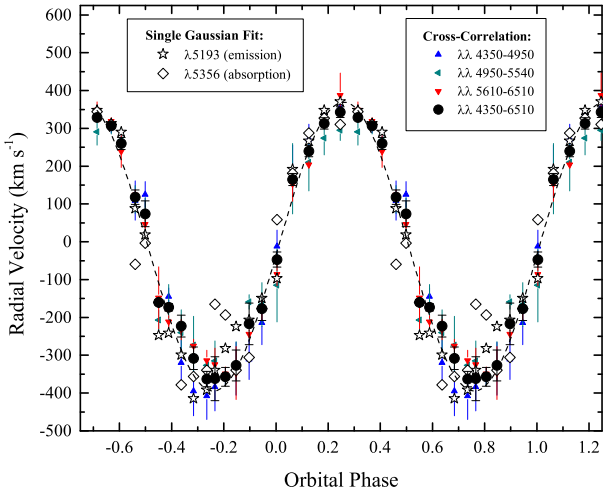}
}
\caption{Radial velocities of the secondary star folded on the ephemeris from Table~\ref{Tab:Syspar}.
         The measurements are obtained using the cross-correlation in different wavelength ranges.
         It also shows the radial velocities of the lines \l5193 \AA\ (in emission) and \l5336 \AA\
         (in absorption) obtained with a single Gaussian fit.
         Two cycles are shown for clarity.
         }
\label{Fig:RadVelSec}
\end{figure}

The first step was to measure the radial velocity variations of the strongest observed lines \l5193\,\AA\
and \l5356. The velocities were independently measured by fitting each line profile in the phase-folded
spectra with a single Gaussian.
The resulting radial velocity curves were then fitted with a sinusoid of the form
  \begin{equation}  \label{radvelfit2}
    V(\phi)=\gamma -K_{2,\rm o} \sin \left[ 2\pi \left(\phi-\phi_0 \right) \right] .
  \end{equation}
We obtained the observed radial velocity semi-amplitude $K_{2, \rm o}$ to be $383\pm13$ \kms\ for
the emission line \l5193\,\AA\ and $349\pm20$ \kms\ for the absorption line \l5356\,\AA, whereas
the difference between the phase zero-points $\phi_0$ for these lines was found to be $\sim$0.01
(Fig.~\ref{Fig:RadVelSec}). Using these preliminary values of $K_{2, \rm o}$ and $\phi_0$,
each individual spectrum was then shifted to correct for the orbital motion of the donor star.
The cross-correlation template was then obtained by averaging shifted individual spectra.

%-----------------------------Figure Start------------------------------
%-----------------------------Figure End--------------------------------

\begin{figure}
\centerline{
  \includegraphics[width=8.0cm]{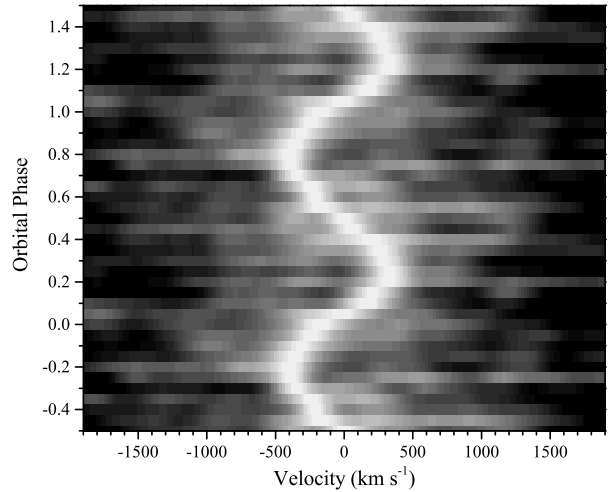}
}
\caption{Two-dimensional representation of the cross-correlation function obtained over the wavelength
         interval of \l\l4350--6510~\AA. Two cycles are shown for clarity.
         }
\label{Fig:CrossCor}
\end{figure}

In the next step, the phase-folded spectra of \source\ were cross-correlated with the template.
Prior to the cross-correlation, the target and template spectra were normalised by dividing by
the result of fitting a low-order spline to the continuum. In a case of noise-dominated spectra,
its cross-correlation with any template may produce spurious peaks in the cross-correlation
function (CCF) and the correspondingly measured radial velocities would not follow a common sine
curve. Therefore, in order to test for reliability in the derived parameters, the cross-correlation
was carried out in three separate wavelength regions
\l\l4350--4950~\AA, \l\l4950--5550~\AA\ and \l\l5610--6510~\AA. To avoid the influence of the
emission and the night-sky lines, the portions of the spectra around these spectral features were
masked. We found that all the corresponding CCFs show strong and distinct peaks
(Fig.~\ref{Fig:CrossCor}).
The radial velocity and its accuracy were then determined by fitting the strongest peak
of each CCF with a Gaussian and a linear background. Our solutions obtained by fitting the measured radial
velocities with the sinusoid (equation \ref{radvelfit2}) are very similar for all three wavelength ranges
and they are close to the measurements obtained for the visually detected lines.

In the final step, these solutions were used to create a new cross-correlation template. The cross-correlation
analysis was then performed over the full wavelength interval of \l\l4350--6510~\AA\ giving
the best-fitting results $K_{2,\rm o}=382\pm8$~\kms and the phase zero-point of
inferior conjunction of the secondary star ($\phi_0=0$) corresponding to $T_0$=HJD~245\,6510.8081$\pm0.0005$.
In Fig.~\ref{Fig:RadVelSec} we show the measured radial velocities together with their
sinusoidal fit.

The similarity of the solutions in different wavelength regions leaves no doubts in the reality of
the radial velocity variability. The presented analysis also strongly indicates that in addition
to the visible spectral lines, there must exist other emission or absorption features in all parts
of the spectrum, which are undetectable by eye, but distinguishable by cross-correlation. Using the
final cross-correlation template, which essentially is the spectrum of \source\ corrected for
orbital motion of the secondary star, we made another attempt to identify the spectral features
of a late-type star, but were again unsuccessful.

%-----------------------------Figure Start------------------------------
%-----------------------------Figure End--------------------------------

A possible reason for our failure may be the extreme conditions under which the donor star existed
at the time of our observations. During the outburst stage the secondary should be significantly heated
by its companion. Theoretical studies show that a dramatic temperature inversion is expected in
the atmosphere of an irradiated M dwarf star. As a result, the emergent spectrum will be
significantly different from that of an isolated M dwarf \citep{Barman}. Depending on the power
of the incident flux and the temperature, pressure,
and chemical composition of the secondary's atmosphere, the resulting spectrum can have complex
mixtures of emission and absorption lines.
The identification of such a spectrum would be a very complicated task.

\subsection{Disc emission lines}
\label{DopMapSec}

We compared the emission lines between the UV and optical data sets and show mean profiles in
Fig.~\ref{Fig:Profiles}.
In contrast to the UV lines, the optical lines are much weaker (e.g. the
peak normalized fluxes of \HeII\ $\lambda$1640 and $\lambda$4686 are 1.37 and 1.08, respectively),
nevertheless their profiles are very similar in shape and width.

The lines are very wide, and the higher excitation lines are much wider than \Halpha.
This property and the double-peaked appearance of the emission lines suggest their origin
in an accretion disc \citep{Smak1981,HorneMarsh}. Table~\ref{LineParamTab} outlines
different parameters of the major emission lines measured from the averaged spectra.

\subsubsection{Accretion disc parameters from modelling of the emission line profiles}
\label{Sec:modelling}

Despite the similarity of the emission line profiles in \source\ with those typically observed
in other LMXBs, the steepness of the profile wings of high excitation lines (\HeII\ and \CIV)
in \source\ is quite unusual. It is well known
that the shape of the double-peaked profile wings is controlled by the surface radial emissivity
profile \citep{Smak1981,HorneMarsh}. This profile is commonly assumed to follow
a power-law model of the form $f(r)\propto r^{-b}$, where $r$ is the radial distance from the
compact object. The mechanism powering the emission lines in quiescent accretion discs
is discussed in \citet{HorneSaar}. Observations of cataclysmic variables (CVs) and BH binaries
show that $b$ is usually in range of 1--2, rarely being less than 1.5 \citep{JKO89,Orosz94,Orosz02}.

In order to estimate $b$ and other parameters of the accretion disc of \source, we fitted the
symmetrical double-peaked emission line profiles using a simple model of a uniform flat
axisymmetric Keplerian geometrically thin disc \citep{Smak1981}. To calculate the line
profiles we used the method of \citet{HorneMarsh}.
Examples of the application of this technique to the real data are given in
\citet{JKO89,Orosz94,Orosz02,NeustroevWZ,NeustroevIP}.
The three primary free parameters of the model are
\begin{enumerate}
  \item $V_{\rm out}$, the velocity of the outer rim of the accretion disc;
  \item $b$, the power-law index of the line emissivity profile $f(r)$;
  \item $r_{\rm in}/r_{\rm out}$, the ratio of the inner to the outer radii of the disc.
\end{enumerate}

\begin{figure}
\begin{center}
\centerline{
  \includegraphics[width=8.5cm]{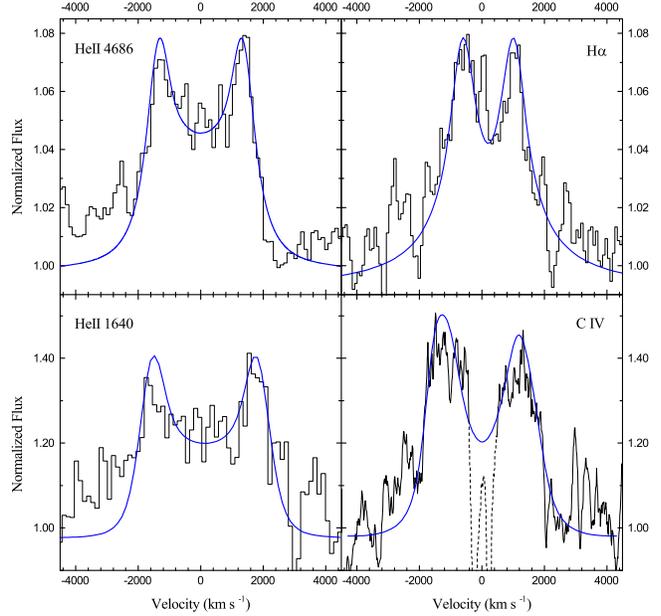}
}
\end{center}
\caption{The profiles of the emission lines observed in the optical and UV spectra of \source\
  together with the corresponding model fits.}
\label{Fig:Profiles}
\end{figure}

Fig.~\ref{Fig:Profiles} shows four optical and UV emission line profiles together with
the corresponding model fits.
Parts of the profiles affected by other spectral features
(such as deep interstellar absorptions in the centre of \CIV\ and a strong emission feature in the
centre of \Halpha) were excluded from the fit.
The best-fitting model parameters are listed in Table~\ref{LineParamTab} and
the errors were estimated with a Monte Carlo approach described in \citet{BorisovNeustroev}.
The best-fitting power-law index for \Halpha\ line was $b=1.58$, which is very
close to the values of $b$ found for many other LMXBs. However, the model fits for \HeII\ and
\CIV\ lines give $b$$\approx$0 and even negative values that suggests an unusually flat (or even
inverted) radial distribution of the emission-line flux from the accretion disc of \source.
Because of the lack of data, it is not clear whether such a behaviour is common for BH binaries.
However, \citet{Marsh-Horne-90} found that in the dwarf nova IP\,Peg during an outburst stage, the \HeII\
\l4686 radial emissivity profile is also remarkably flat compared to the Balmer emission ($b\approx$0.1
and $\approx$2, respectively). \citeauthor{Marsh-Horne-90} showed that such a behaviour is not in
agreement with the predictions of line emission from optically thin discs. They concluded that
photoionization by the soft X-rays and UV photons generated in the centre of the accretion disc should be
taken into account. We suggest that a similar mechanism may explain the observed properties of the
emission lines of \source.

\begin{table*}
\begin{center}
\caption{Parameters of the major emission lines in the averaged spectrum of \source.}
\label{LineParamTab}
\begin{tabular}{lclccccrc}
\hline
Spectral       &  Flux            &   EW    & Relative  & FWHM   & Peak-to-peak  & \multicolumn{3}{c}{Model parameters}\\
line      & (\t $10^{-14}$ erg\,s$^{-1}$cm$^{-2}$)    & (\AA)   & flux & (\kms) & (\kms)        & $V_{\rm out}$ (\kms) &  $b\qquad$  &  $r_{\rm in}/r_{\rm out}$\\
\hline
\Halpha        &  0.223 &  3.6   & 1.07      & 2450    & 1650          &  798$\pm$10 &  1.58$\pm$0.05 & 0.02$\pm$0.01 \\
\HeII\ \l4686  &  0.329 &  4.3   & 1.08      & 4200    & 2690          & 1370$\pm$6 & $-$0.80$\pm$0.11 & 0.06$\pm$0.05 \\
\HeII\ \l1640  &  1.21 &  7.1   & 1.37      & 4020    & 3400          & 1677$\pm$7 & $-$0.85$\pm$0.11 & 0.14$\pm$0.06 \\
\CIV\ \l1548.2 &  \multirow{2}{*}{1.33} & \multirow{2}{*}{7.1} & \multirow{2}{*}{1.45} & \multirow{2}{*}{3820} & \multirow{2}{*}{2440} & 1348$\pm$8 & 0.05$\pm$0.12 & 0.14$\pm$0.06 \\
\CIV\ \l1550.8 &                  &        &           &         &               & 1352$\pm$9 & 0.02$\pm$0.15 & 0.10$\pm$0.04 \\
\hline
\end{tabular}
\end{center}
\end{table*}

%-----------------------------Figure Start------------------------------
%-----------------------------Figure End--------------------------------

%-----------------------------Figure Start--------------------------------

\begin{figure*}
%\begin{center}
\hspace{0.6cm}
\includegraphics[width=4.0cm]{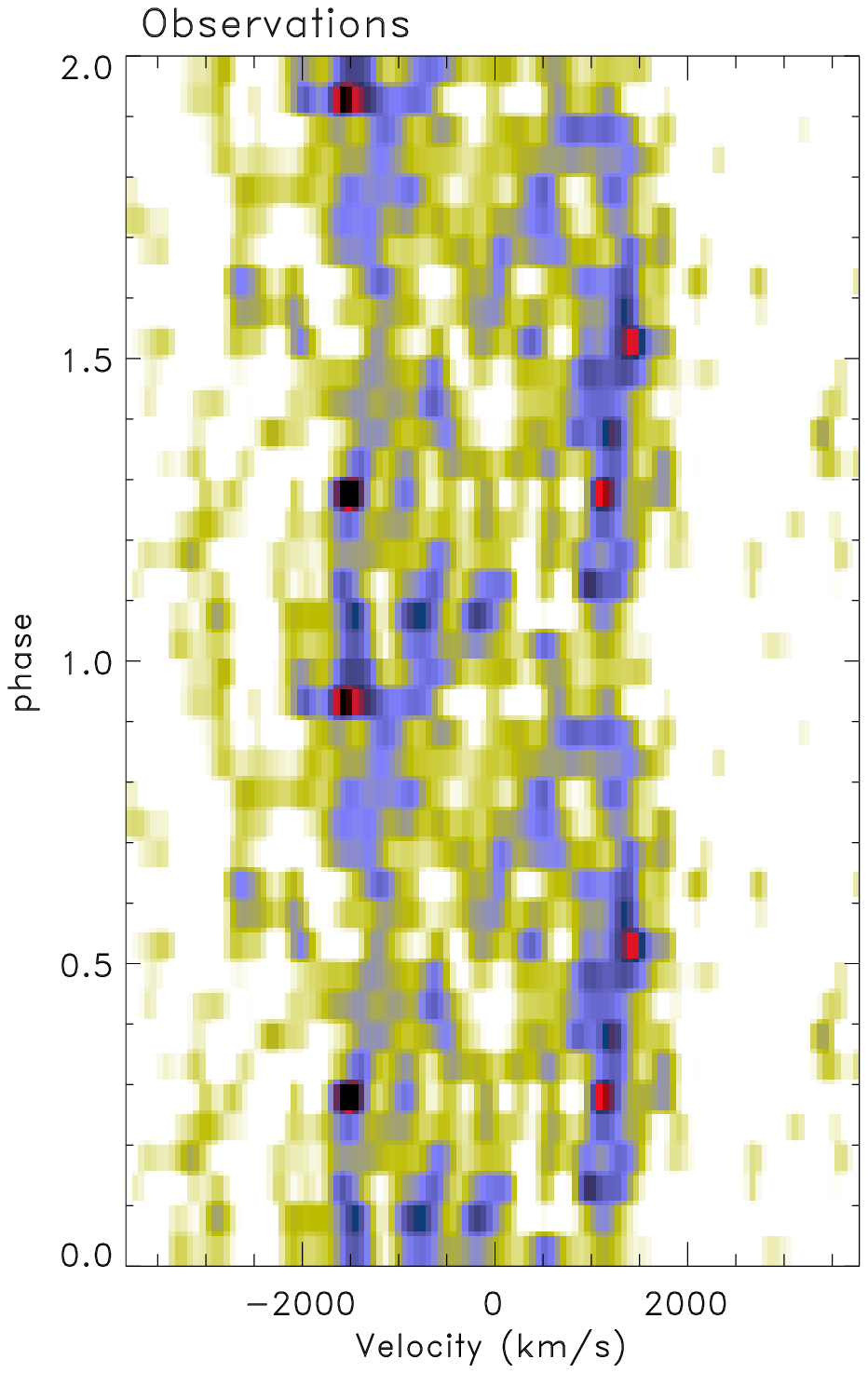}
\includegraphics[width=4.0cm]{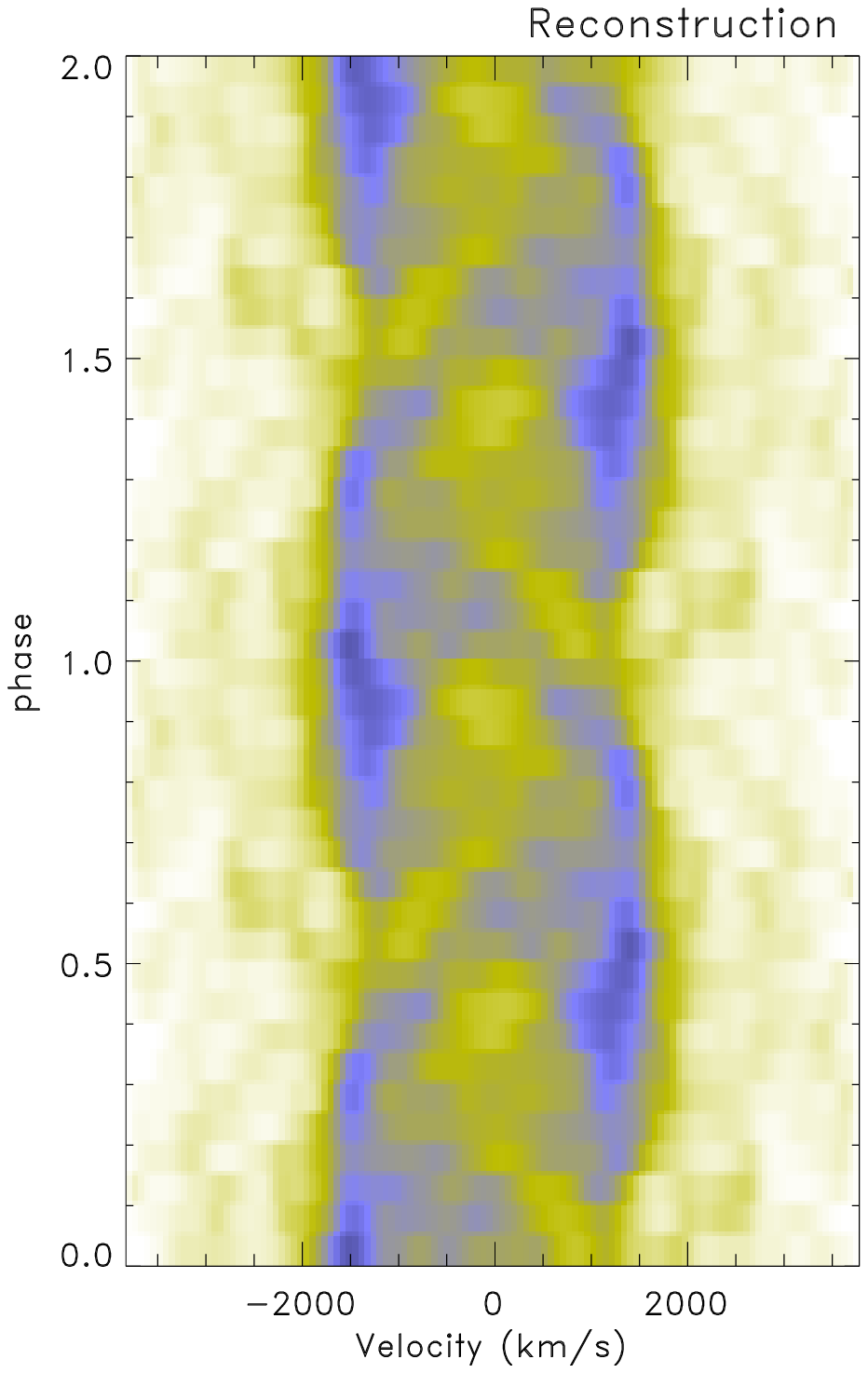}
\hspace{0.4cm}
\includegraphics[width=4.0cm]{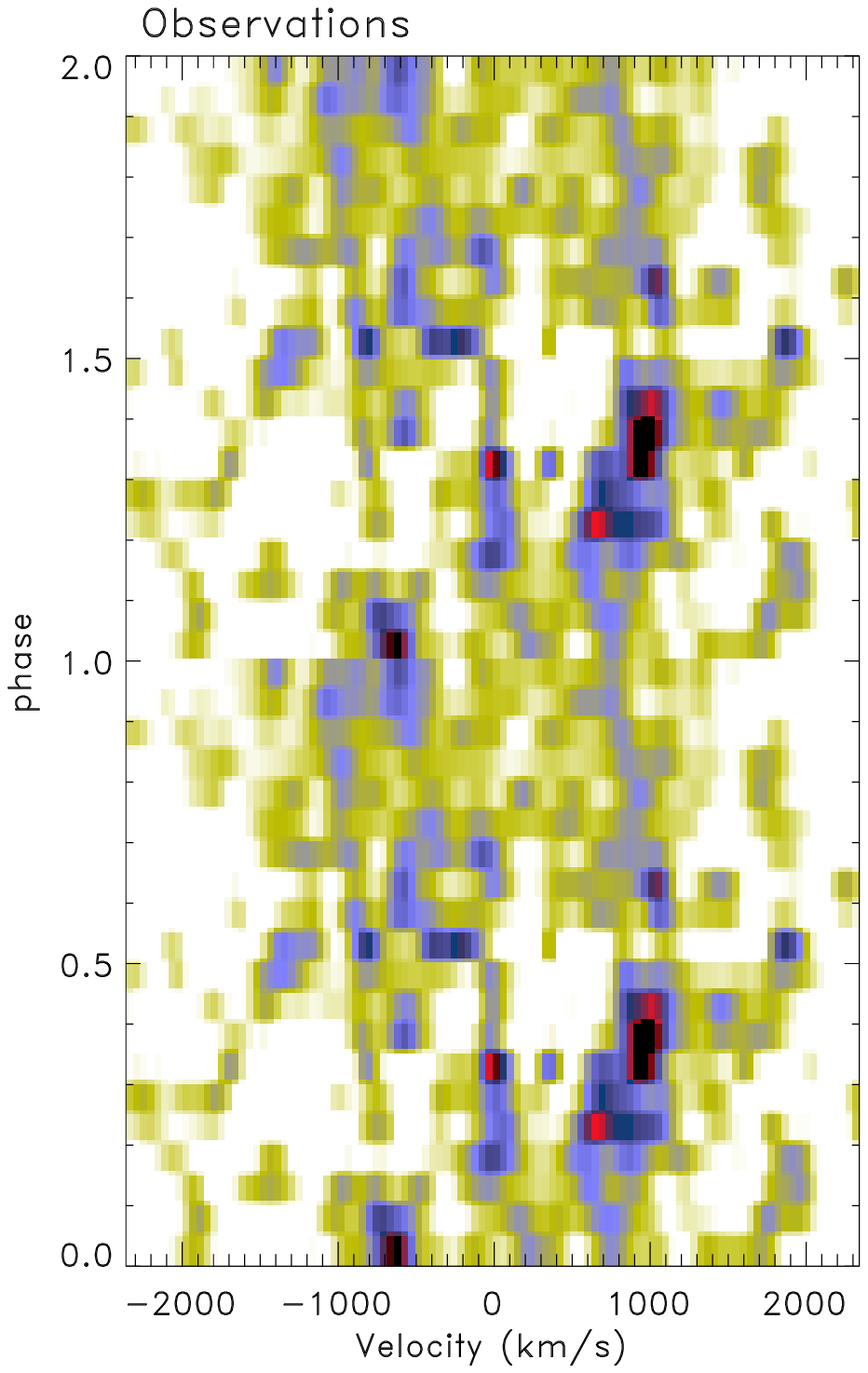}
\includegraphics[width=4.0cm]{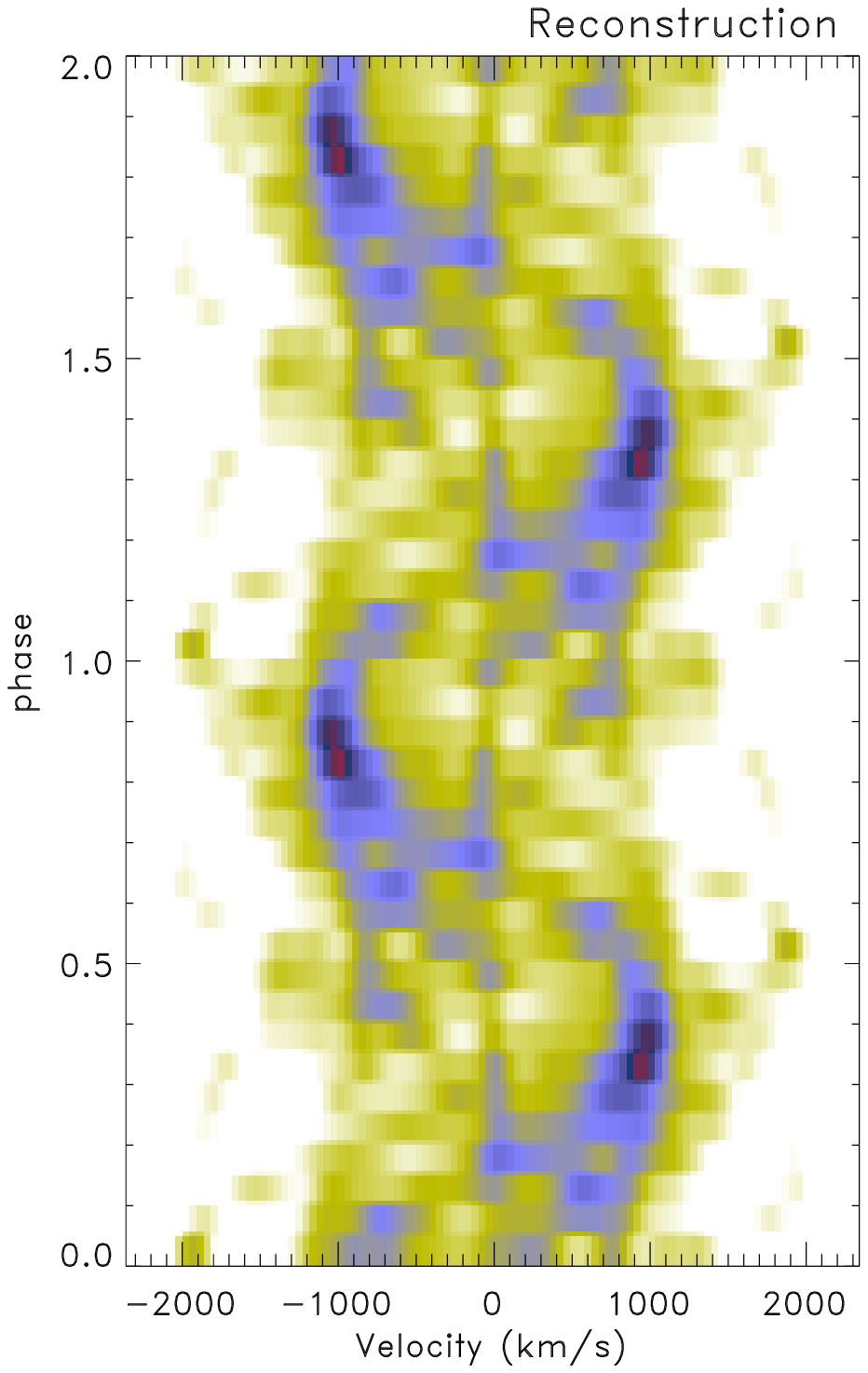}\\
\includegraphics[width=8.5cm]{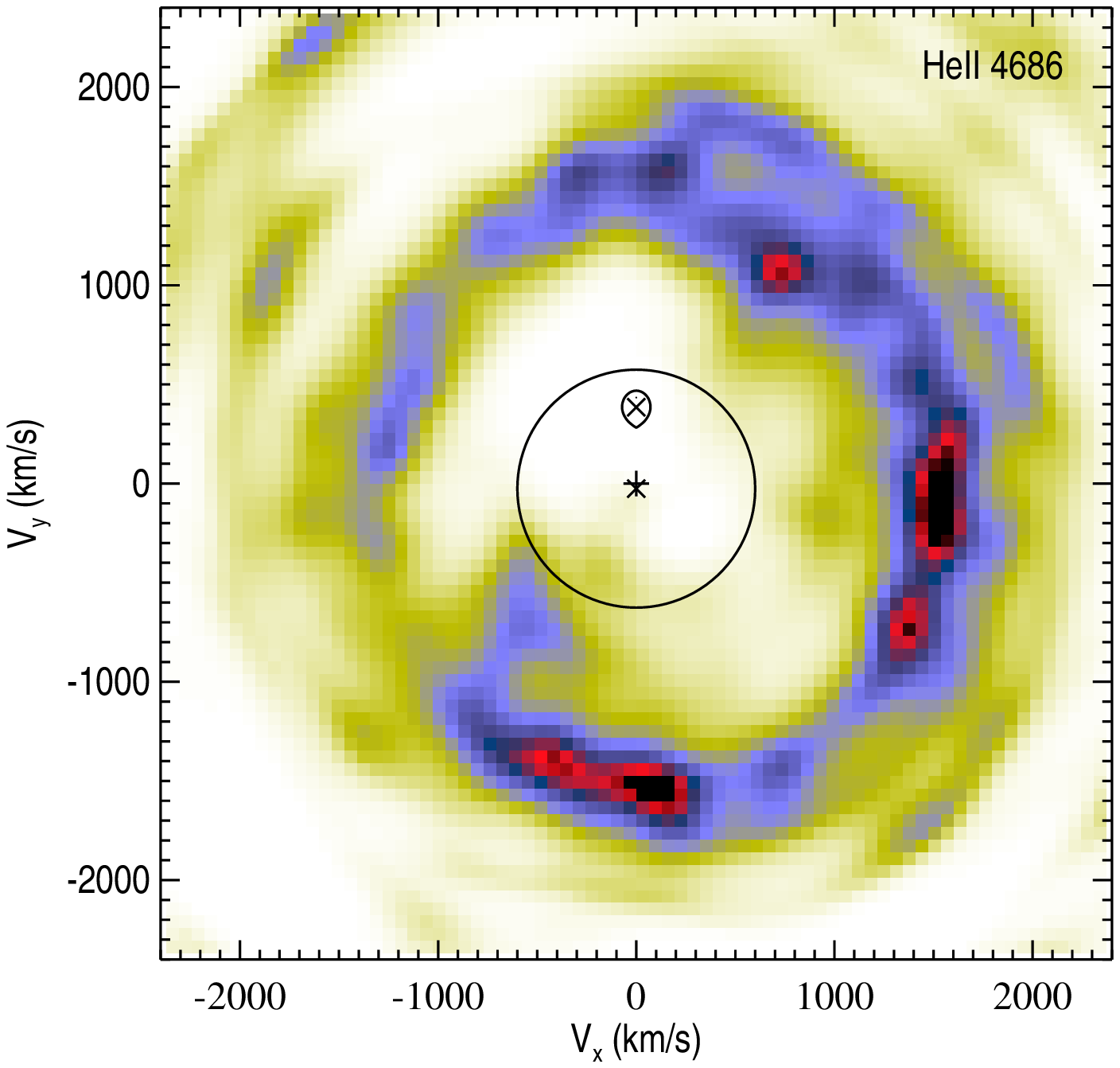}
\includegraphics[width=8.5cm]{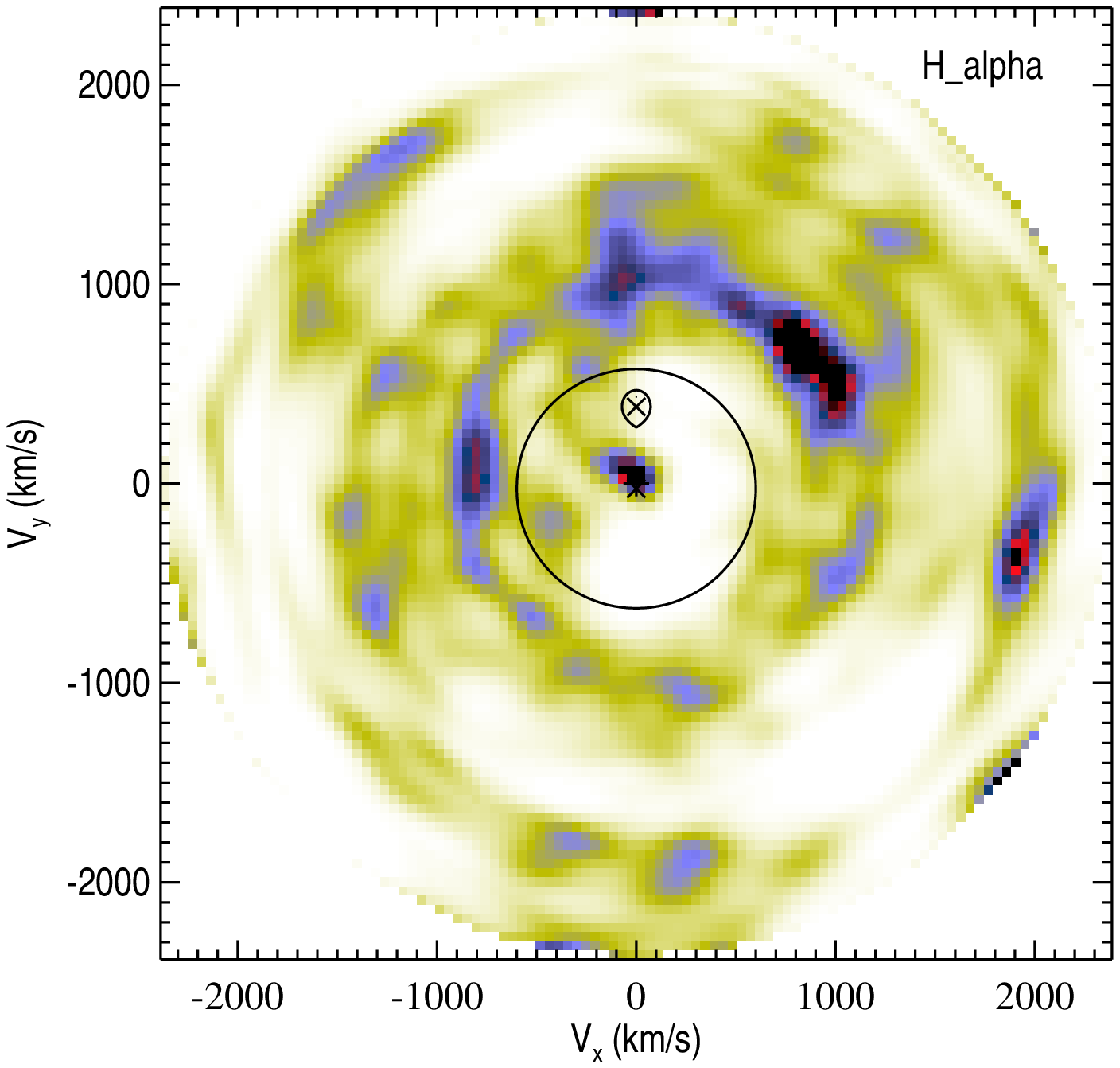}
%\end{center}
\caption{Doppler tomography for the \HeII\ $\lambda$4686 and \Halpha\ emission lines.
The top panels show the observed (left) and reconstructed (right) line profiles
folded on the orbital period. The corresponding
Doppler maps are shown in the bottom panels.}
\label{Fig:dopmap}
\end{figure*}
%-----------------------------Figure End--------------------------------

\subsubsection{Orbital variability of the optical emission lines, the equivalent widths and
               Doppler tomography}

\begin{figure}
\centerline{\includegraphics[height=6.5cm]{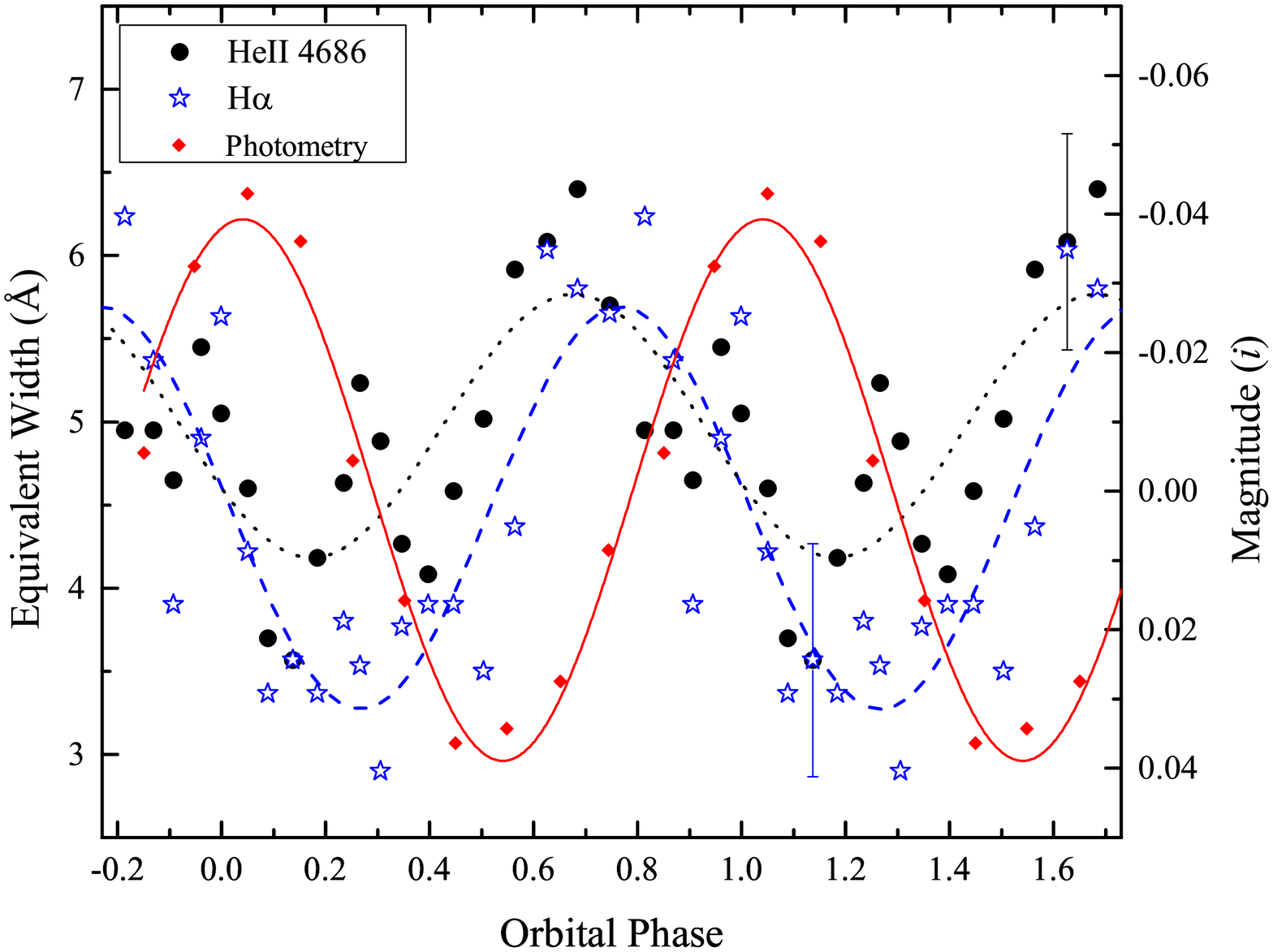}}
\caption{The variation of the EWs of \HeII\ $\lambda$4686\ (filled circles) and \Halpha\
(open stars) with the orbital period. Also shown the phase-binned $i$-band light curve (red diamonds,
the right-hand axis). Two periods are shown for clarity.}
\label{Fig:EW}
\end{figure}

A visual inspection of the optical spectra has shown  orbital variations in
the line profiles. This is better seen in trailed spectra  as variations of the relative intensity
of the two peaks in the double-peaked profiles (see top
left panels of Fig.~\ref{Fig:dopmap}).
We measured the EWs of the emission lines in the phase-folded spectra using
the {\sc iraf} task \emph{splot} (Fig.~\ref{Fig:EW}). The 1$\sigma$ errors
were estimated from the noise fluctuations in the  continuum. Though the obtained
EWs exhibit rather significant dispersion, none the less we see that they are modulated
with the amplitudes of about $\pm$25\% of the mean value.

The observed EW minima could be due to an increase in the continuum luminosity when the region
of enhanced emission crosses the line-of-sight.
There is, however, an apparent shift between the EW
curves and optical light curves (see the phase-binned $i$-band light curve in
Fig.~\ref{Fig:EW}). The maximum of the optical flux occurs close to the inferior conjunction
of the secondary star at phase 0, whereas the EW minima are observed around phase 0.2.

The orbital variation of the emission line profiles indicates the presence of a non-uniform
structure in the accretion disc. In order to study it in more detail, we used Doppler tomography.
Technical details and examples of the application of Doppler tomography to real data are given by
\citet{Marsh-Horne-88} and \citet{marsh2001}. The Doppler maps of the \HeII~$\lambda$4686
and \Halpha\ emission lines were computed using the code developed by \citet{Spruit}. The
resulting tomograms are presented in the bottom panels of Fig.~\ref{Fig:dopmap}, whereas
the observed and reconstructed emission line profiles are shown in the top panels.
To help in interpreting the Doppler maps, additional symbols are inserted, which mark the positions
of the compact object (lower cross), the centre of mass of the binary (middle cross) and the Roche
lobe of the secondary star (upper bubble with the cross). The circle of radius 600 \kms\ around
the center of mass corresponds to the projected velocity of the outer part of the largest accretion
disc restricted by tidal forces. The Roche lobe of the secondary has been plotted using the system
parameters: $M_1$=3\Msun, $M_2$=0.2\Msun, $i$=40\degr (see Section~\ref{Sec:SysPar}).

The observed sharp asymmetric double-peaked profiles of the emission lines produced
the azimuthally asymmetric annuli of emission. The radii of the annuli are different
for \HeII\ and \Halpha, reflecting the
different peak-to-peak velocity separation in these lines. However, the overall appearance
of the tomograms is rather similar. They both display the enhanced emission region in the
upper-right quadrant. The origin of this structure is unclear.
It is located far from the region of interaction between the stream and the disc, which is
situated to the left of the secondary star bubble. The data neither show the emission from
the donor which is often observed in the CVs during outbursts\footnote{However, a bright
compact spot at the position of the secondary star is clearly seen in a Doppler map of the
emission line at \l5193\AA.}.

We also note that the \Halpha\ data display the low-velocity component which produces a compact
spot of emission located around the centre of mass of the binary in the corresponding Doppler map.
The source of this feature is not clear. It might be a result of poor subtraction of the geocoronal
\Halpha\ night sky line. However, the phase dependence of this line raise a doubt.

%-----------------------------Figure Start------------------------------
%-----------------------------Figure End--------------------------------

\subsubsection{Radial velocity analysis}

We attempted to derive the radial velocity curve of the compact object in \source\ using the
double-Gaussian method \citep{sch:young,Shafter}, but results were implausible. This is not
a surprise, as the method uses the extreme wings of the emission line profile which are very
noisy even in the phase-folded spectra of the object because of their steepness.
Instead, the radial velocities were measured by fitting the modelled
double-peaked profile (see details in Section~\ref{Sec:modelling}) to the observed phase-binned profiles. All the primary parameters of the
model ($V_{\rm out}$, $b$ and $r_{\rm in}/r_{\rm out}$) were frozen to the averaged values listed in Table~\ref{LineParamTab},
but the model profile was allowed to shift along wavelengths. We used only the \HeII\ $\lambda$4686 line
for this analysis as it has more clear and symmetric profiles than \Halpha\ and also because it is presumably
formed in the inner parts of the accretion disc. Therefore, it should represent the motion of
the central star with a higher reliability.

\begin{figure}
\centerline{
  \includegraphics[height=6.5cm]{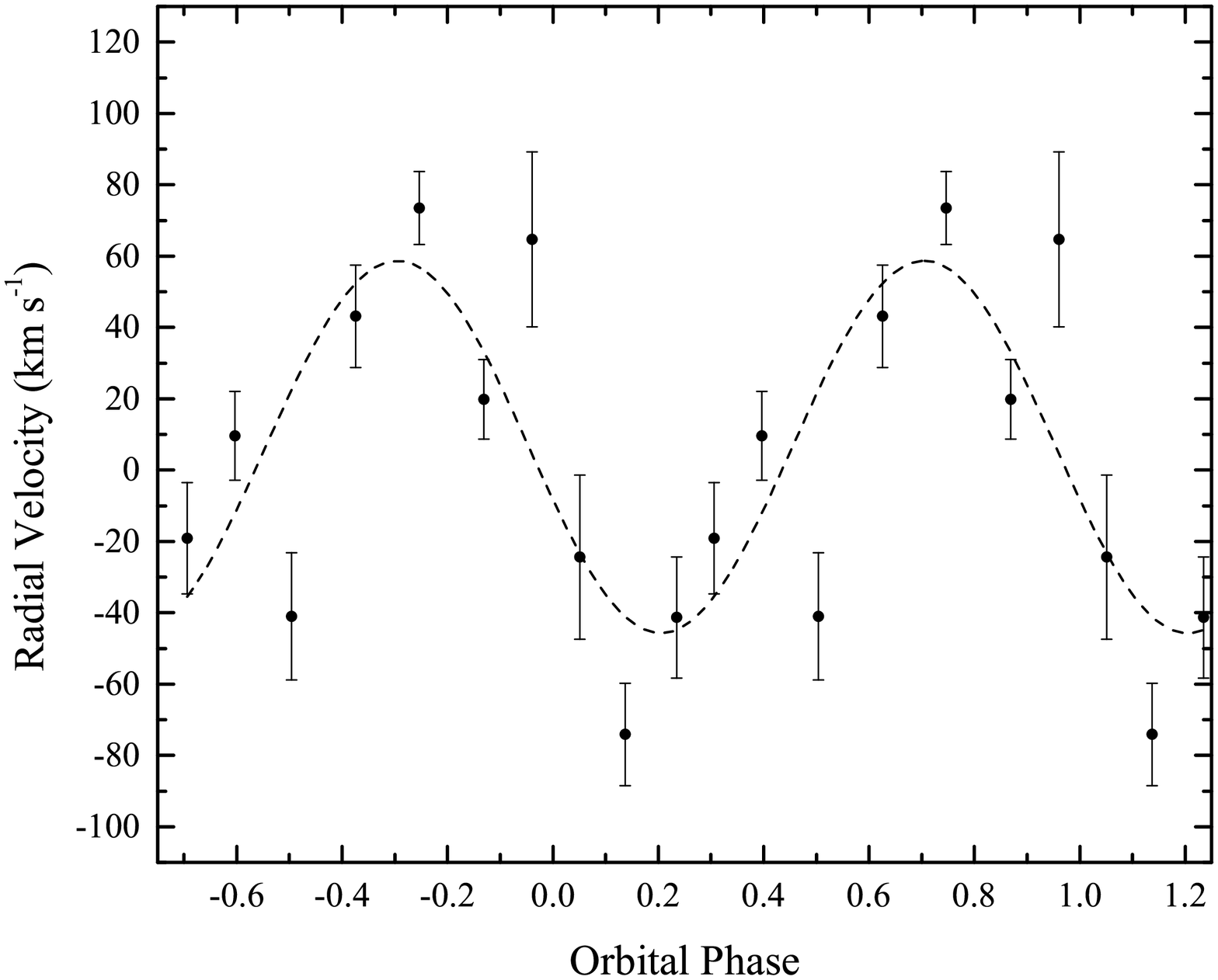}
}
\caption{Radial velocities of the \HeII\ $\lambda$4686 emission line folded on the ephemeris from
Table~\ref{Tab:Syspar}. Two cycles are shown for clarity.
         }
\label{Fig:RadVel}
\end{figure}

The results are shown in Fig.~\ref{Fig:RadVel} (right-hand panel). The errors are the formal model fitting
errors estimated with a Monte Carlo approach. We made a non-linear least-square fit of the derived
velocities to a sinusoid of the form (equation \ref{radvelfit2}), where $K_2$ was replaced by $K_1$.
This fit gives the semi-amplitude $K_1=52\pm10$ \kms, the systemic velocity $\gamma=6\pm6$ \kms\
and $\phi_0=0.45\pm0.03$. The difference between the phase zero-points obtained from
the emission and absorption lines is close to 0.5, as expected if the derived velocities from those
lines trace the motion of the two components. However, it is well known that the parameters obtained
with this method are affected by systematic errors, because the emission lines arising from the accretion
disc may have severe asymmetric distortions (see, for example, discussion in \citealt{Orosz94}).
Thus, the obtained value of $K_1$ should be used with great caution.

%-----------------------------Figure Start------------------------------
%-----------------------------Figure End--------------------------------

\section{The binary system parameters}
\label{Sec:SysPar}

\subsection{The $K$-correction}

The observed values of $K_{2,\rm o}$ are often known to suffer from non-uniform distribution of
absorption or emission spectral features on the donor star surface due to the heating effect by
irradiation from the X-ray source. The noncoincidence of the centre-of-mass and the centre-of-light
of the donor can result in systematic errors in the determined velocity amplitude $K_{2,\rm o}$.
A mixture of emission and absorption lines in the spectrum of \source\ strongly suggests that
irradiation of the secondary plays a significant role.
Moreover, the flux at 5000 \AA\ from a low-mass star with
temperature of 3000 K and radius of 2\tim{10} cm at 2 kpc is expected to be considerably smaller
than the observed one ($F_{\l}\approx$ 2.5\tim{-19} and 7\tim{-16} \Ergs, respectively). Thus,
no lines from the dark side of the star should be visible in the observed spectrum. This
suggests that the cross-correlation signal likely originates in the irradiated surface of the
secondary. Therefore, in order to obtain the true value of $K_2$, a `$K$-correction' should be
applied.

The $K$-correction generally depends on the mass ratio and the distribution of spectral line
emission or absorption over the surface of the secondary star and can be expressed as
(\citealt{WadeHorne}; see also equation 2.77 in \citealt{Warner}):
\begin{equation}
\label{Eqn:Kcor}
 \frac{K_{2,\rm o}}{K_2} \approx 1-0.462 q^{1/3} (1+q)^{2/3} \frac{\Delta R_{2}}{R_{2}} \,,
\end{equation}
where $q$ is the mass ratio $M_2/M_1$, $R_{2}$ is the radius of the secondary and $\Delta R_{2}$
is the displacement of the centre-of-light from the centre of mass of the secondary
star. It is difficult to quantify the correction without the knowledge of these parameters. In
order to estimate its reliable extreme value, we adopted the realistic values of $q=0.04-0.12$
(see Section~\ref{sect:montecarlo}) and assumed $\Delta R_{2}/R_{2}$ to be $\sim$0.5 that roughly
corresponds to the extreme case where the spectral lines come only from the hemisphere closest to
the primary. Within these assumptions $K_{2,\rm o}/K_2$ is in the range 0.88--0.92, therefore in the
following analysis we use $K_{2,\rm o}/K_2=0.9$ as an illustration for a possible $K$-correction.

In theory, the correction can be larger if emission is tightly concentrated near the L$_1$-point.
However, this is not supported by observations. Such a compact emission source should be luminous
enough to produce a visible structure in the \Halpha\ Doppler map. Moreover, in a realistic situation,
the accretion disc's outer rim casts a broad shadow on to the donor star, and only relatively small
areas near the poles of the donor's facing hemisphere can be directly irradiated by the central X-ray
source. In order for the secondary star to be illuminated, indirect irradiation via scattered X-rays
should be taken into account, or there should exist another source of irradiation, located outside
the equatorial plane. Both these scenarios imply that sufficiently extended regions on the donor star
will be irradiated (for a detailed discussion, see \citealt{Dubus}; \citealt{Ritter}).

\subsection{Mass function and constraints on the compact object mass from the double-peaked emission lines}
\label{Sec:DoublePeakedConstraints}

The  semi-amplitude $K_2$ of the secondary's radial velocity curve and the orbital period $P_{\rm orb}$
give the mass function which sets an absolute lower limit for the mass of the compact object:
\begin{equation} \label{MassFunc}
   f(M) = {K_2^3 P_{\rm orb} \over 2 \pi G} = {M_1^3 \sin^3 i \over (M_1 + M_2)^2} ,
\end{equation}
where $M_1$ is the mass of the compact object , $M_2$ is the mass of the secondary star and $i$
is the binary inclination. From the observed velocity $K_{2,\rm o}=382$ \kms\ we can calculate the
``observed'' mass function $f_{\rm o}(M) = 0.69\pm0.04$ \Msun. Applying the $K$-correction, we obtain
the mass function $f(M) \lesssim 0.95$ \Msun, which is one of the smallest mass function for a BH LMXB
\citep{CasaresJonker}.

Another independent approach to constrain the compact object mass is to measure the outer disc velocity.
\citet{Smak1981} has shown that assuming the Keplerian velocity in the accretion disc
\begin{equation}
\label{VelKep}
  V_{\rm K} = \sqrt{{G M_1 \over r}}\,,
\end{equation}
the double-peaked profiles can be used to determine the projected velocity $V_{\rm out}$ of the outer
rim of the accretion disc, which, in turn, depends on the mass $M_1$ of the accreting star and the
radius of the disc. The largest radius of the accretion disc, should it be determined from
observations, can further constrain the system parameters.

We measured $V_{\rm out}\approx800$~\kms\ through the modelling of the \Halpha\ emission line, which
originates in the outermost part of the accretion disc (see Section~\ref{Sec:modelling} and
Table~\ref{LineParamTab}). We note, however,
that during our observations both \Halpha\ and \HeII\ \l4686\ lines were much wider than they were
at the beginning of the outburst, when the peak-to-peak separation for \Halpha\ was 1200--1300~\kms\
\citep{Torres05b}.

This implies that the accretion disc of \source\ has shrunk by about 45 per cent
compared to the early phase of the outburst,
similar to what was observed in other LMXBs \citep{XTEJ1118} and in CVs \citep{Warner}.
In this context we note that the outer parts of a large accretion disc are under the gravitational
influence of the secondary star, which prevents the disc from growing above the tidal truncation
radius $r_{\rm max}$, where the tidal  and viscous stresses are comparable (\citealt{Warner} and
references therein). It can be estimated as (see equation 2.61 in \citealt{Warner}):

\begin{equation}
\label{RmaxA}
   r_{\rm max}  =a  {0.6 \over 1+q}\,,
\end{equation}
where $a$ is the binary separation.
Combining Kepler's third law with equations ~(\ref{VelKep}) and (\ref{RmaxA}), we obtain
the relation:
\begin{equation} \label{MassSum}
   (M_1 + M_2)\,\sin^3 i = {0.074\,P_{\rm orb} V_{\rm K, max}^3 \over G}\,,
\end{equation}
where $V_{\rm K, max}$ is the projected Keplerian velocity of the accretion disc at the tidal radius.
Adopting Torres et al.'s value for $V_{\rm out}$ of 600~\kms\ as an estimate of $V_{\rm K, max}$
we get
$(M_1 + M_2)\,\sin^3 i = 1.2$ \Msun.
Using the realistic value of 0.2 \Msun\ for the secondary mass (see Section~\ref{Sect:comp_mass}),
we find that the solutions for $M_1$ for any inclination are in good agreement with the solutions
obtained from the mass function with the $K$-correction applied (Fig.~\ref{Fig:mass}).

\subsection{Constraints on the secondary mass}
\label{Sect:comp_mass}

It is clear that the secondary is a low-mass star, otherwise its absorption lines would be much more
apparent in the spectrum of \source.
In order to be transferring matter on to the compact component, the secondary star must fill its Roche lobe.
The relative size of the donor star is therefore constrained by the Roche geometry and the donor must obey
the period-density relation for the Roche lobe filling objects \citep{Warner}.
If the secondary is a main-sequence star, one immediately gets an estimate for the mass. The
empirical and theoretical mass-period relations for a 2.85~h orbital period binary yield the
mass of the secondary star in the range 0.17--0.25 \Msun\ \citep{Warner,Smith:Dhillon,Patterson05}.
If the donor is an evolved star, it is likely somewhat inflated
relative to isolated main-sequence stars of the same mass, thus the mass inferred from such relation
can be considered as an upper limit \citep{KKB01,Knigge_rev}.
The recently calculated evolutionary sequences for NS binaries with a 3~h period obtain
the secondary masses in the range 0.1--0.3 \Msun \citep{PRP02,LRP11}.
We conservatively assume this mass range for the secondary in \source.

\subsection{Constraints on the inclination}
\label{Sec:Inclination}

It is clear from the discussion above that the companion mass is rather low, and therefore
the BH mass mostly depends on the unknown inclination of the system and the possible $K$-correction.
Even though the BH mass can be rather large for small $i$, there are additional observational
reasons which argue against a very low orbital inclination angle.
\begin{enumerate}
 \item The UV spectrum of \source\ is dominated by relatively strong emission lines \CIV\ and
       \HeII\ without P~Cyg absorption components, and no other absorption lines are detected.
       From studies of CVs with high accretion mass rates (nova-like stars and dwarf novae in
       outbursts) it is known that the appearance of their UV spectra strongly depends on
       orbital inclination: low inclination systems show mainly absorption features in spectra,
       intermediate-to-low inclination systems exhibit P~Cyg profiles and/or blue-shifted deep
       absorptions, whereas high inclination systems show strong emission lines
       \citep{LaDous,Puebla2007,Puebla2011}. It is not clear if these results can be applied to
       LMXBs whose accretion discs are strongly irradiated by the X-rays from the inner region.
       Published UV spectra of LMXBs show relatively strong emission lines. However,
       to the best of our knowledge, no UV observations of any low or intermediate-to-low
       inclination systems are available in the literature. Thus, based on the results of the
       numerical simulations of CVs \citep{LaDous,Puebla2011}, we can put a lower limit of
       $i\ga40\degr$.
 \item The time resolution optical light curves of \source\ obtained during our observations
       display prominent orbital variations with the amplitude of 0.10--0.15 mag in $V$
       (Fig.~\ref{Fig:Ratir}), whose
       nature is not yet clear. The source of orbital modulation cannot be the varying aspect
       of the irradiated secondary star because the maximum of light is observed around phase
       0.0, i.e. inferior conjunction of the secondary star. This also can hardly be a bright
       spot or bulge on the outer edge of the disc formed by the impact of the gas stream. \source\
       is still in outburst, thus its accretion disc is in the hot, ionized state. The observations
       of CVs and LMXBs show that the bright spot is barely seen during the outburst stage. Most
       likely, this orbital modulation is due to some substructures in the non-uniform or/and
       non-axisymmetric accretion disc (i.e. spiral waves, warps or tidally thickened regions).
       The detection of the enhanced emission regions in the Doppler maps and of the EW
       variability supports this idea. It is quite obvious that a low-inclination binary system
       is unable to produce a relatively large amplitude EW variability and orbital modulation
       in the light curve.
       Our simulations have shown (see Appendix~\ref{Appendix}) that for more or less realistic
       parameters of the bright area in the accretion disc it is virtually impossible to obtain
       the orbital variability with the required amplitude for the orbital inclination
       $i\lesssim$40\degr.
\end{enumerate}

%************************  SysPar  ************************************
\begin{figure}
\centering
\includegraphics[width=8cm]{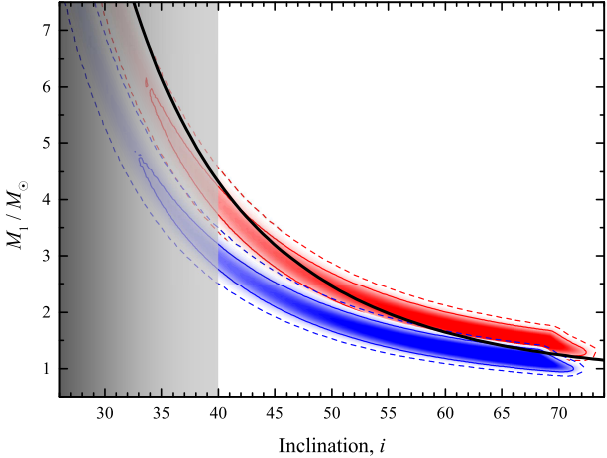}
\caption{
Constraints on the black hole mass $M_1$ in \source\ obtained using a Monte Carlo simulation of
the observed parameters. The blue area corresponds to the observed semi-amplitude $K_{2, \rm o}$,
whereas the red data are calculated for the $K$-correction $K_{2,\rm o}/K_2=0.9$. Denser colours
represent higher probability. The solid and dashed lines are the 68~per~cent and 95~per~cent
confidence levels, respectively. The black line shows the BH mass as set by the
relation~(\ref{MassSum}) for $M_2=$0.2\Msun. The shaded area marks the improbable solutions
because of the observed strong photometric and spectroscopic orbital variability (see the text for
explanation).
}
\label{Fig:mass}
\end{figure}
%************************  SysPar  ************************************

On the other hand, large inclinations are ruled out too, because both ours
and the published photometry and spectroscopy of \source\ are extensive enough to rule out any significant eclipse.
The maximal possible inclination in absence of eclipses generally depends on the mass ratio and can be estimated as
\begin{equation} \label{eq:imax}
 \cos i \gtrsim \frac{R_2/a}{1-r_{\rm d}/a} ,
\end{equation}
where $r_{\rm d}$ is the disc size, which we assume to be equal to 60 per cent of the maximal tidally-allowed size
$r_{\rm max}$ (see Section~\ref{Sec:DoublePeakedConstraints}),
and the nominator is computed from the \citet{Eggleton83} formula:
\begin{equation} \label{eq:r2a}
\frac{R_2}{a} = \frac{0.49 q^{2/3}}{0.6q^{2/3} + \ln (1+q^{1/3}) } .
\end{equation}

\subsection{Observational and Monte-Carlo constraints on the binary system parameters}
\label{sect:montecarlo}

From the above analysis we can put a conservative lower limit for the orbital inclination to be
$i$$\ga$40\degr, constrain its upper limit using Equation~(\ref{eq:imax}), and consequently restrict
the BH mass $M_1$ to the ranges 1.0--3.1 \Msun\ for the observed $K_{2,\rm o}$ and
1.3--4.1 \Msun\ with the $K$-correction applied.

In order to illustrate the cumulative effect of uncertainty in input parameters, we also
applied a Monte-Carlo approach using $10^8$ trials for Gaussian distribution of $K_{2}$ (with
$\sigma=8$~\kms), a top-hat distribution of secondary mass in the limits of 0.1--0.3 \Msun\ and a
uniform distribution of $\cos i$. The simulations were performed for the observed $K_{2,\rm o}$ and
with the $K$-correction applied ($K_2=K_{2,\rm o}/0.9$). For any set ($M_2$, $i$, $K_{2,\rm o}$) we
calculate $M_1$ from the mass function, and the parameters are accepted if there are no eclipses
(equation~\ref{eq:imax}).

Fig.~\ref{Fig:mass} shows the results of the Monte-Carlo simulations with the formal 68 and
95~per~cent confidence levels\footnote{The simulations were performed without additional
limitations which should be applied but which values are not well defined (e.g., the minimal
orbital inclination). This biases the Monte-Carlo tests and may affect the confidence levels.}.
Denser colours represent higher probability reflecting a strong
tendency for lower masses. In the figure we also show the BH mass as set by the relation~(\ref{MassSum})
based on constraints obtained from the double-peaked emission lines. These calculations show that
the compact object mass in excess of 5 \Msun\ is rather improbable,
because it requires an extremely large $K$-correction or a lower inclination angle, which is also
unlikely because of the observed strong photometric and spectroscopic orbital variability.

%************************  SysPar  ************************************
\begin{table}
\centering
\caption{Orbital and system parameters for \source.}
\label{Tab:Syspar}
\begin{tabular}{ll}
\hline
\hline\noalign{\smallskip}
Parameter                           & Value                 \\
\hline
\multicolumn{2}{c}{Observed} \\
$P_{\rm orb}$ (h)                   &  2.85$\pm$0.01         \\
$T_{0}$ (+2450000)                  &  6510.8081$\pm$0.0005  \\
$K_1$ (\kms)                        &  52$\pm10$             \\
$K_{\rm 2, \rm o}$ (\kms)           &  382$\pm$8             \\
$\gamma$ (\kms)                     &  6$\pm$6               \\
$f_{\rm o} (M) /\msun$              & 0.69$\pm$0.04          \\
\multicolumn{2}{c}{Constrained} \\
$i$                                 &  $\ga$40\degr\          \\
$M_{1}/\msun$                       &  $\la$3.1 (4.1)$^a$    \\
$M_{2}/\msun$                       &  0.1--0.3              \\
$q=M_{2}/M_{1}$                     &  $\ga$0.04 (0.03)      \\
$a$/$R_{\odot}$                     &  $\la$1.53 (1.67)      \\
\hline
\end{tabular}
\begin{flushleft}{
$^{a}${The numbers in parentheses correspond to the solutions with the $K$-correction applied.}
}\end{flushleft}
\end{table}

We note that even with the lowest possible upper limit on the inclination and the large $K$-correction,
the solution allows for the compact object mass to be below 1.3 \Msun, that would correspond to a low-mass NS.
Nevertheless, X-ray properties of \source\ are typical for a BH binary and argue strongly in favour
of a BH accretor in the system. This fact can be used to impose an additional restriction on the
solution by requiring that the BH has a mass greater than some specified limit, e.g. the maximum
possible NS mass. Following the modern theoretical calculations, which give the maximum NS mass
$\lesssim$2.4 \Msun\ \citep{Lat12}, this limit can be put at 2.5 \Msun. Nevertheless, admitting that
the value of this limit is not well defined, we leave this discussion for the interested reader to
complete.

The values of the measured and constrained system parameters are summarised in Table~\ref{Tab:Syspar}.
In Fig.~\ref{Fig:Model} we show a schematic representation of the suggested geometry for \source,
plotted using the system parameters: $M_1$=3 \Msun, $M_2$=0.2 \Msun, $i$=40\degr.

The presented BH mass calculations were done with the orbital period of 2.85~h.
However, even if the previous value $\sim$3.2~h found by \citet{Zurita08} appears to be
the correct orbital period, the results will not change significantly due to the relatively
weak dependence of the mass function on the period and a large uncertainty in other parameters.

\section{Discussion}
\label{sect:discussion}

\subsection{The BH mass}

According to the current convention, the black holes are compact objects, whose measured masses exceed the limit of 3 \Msun.
To date, many BH binaries were identified according to the high measured mass function $f(M)>$3 \Msun, which puts an absolute lower limit
for the primary mass \citep{CasaresJonker}.
In a number of other systems, additional constraints were required to obtain the mass of the compact object in excess of this limit.
The inferred masses of BHs used to be well above the value 3 \Msun, while the measured NS masses tend to cluster at $\sim$1.4--1.5 \Msun\
\citep{RM06,Lat12}.
The observed distribution of compact objects thus seems to have a double-peak structure at $\sim$1.5 \Msun\ and $\sim$7 \Msun\ for NSs and
BHs, correspondingly \citep{Bailyn98,OzelBH}, with a gap in the mass range 2--5\Msun.
Investigations of the minimal possible black hole mass \citep{BH-Farr} gave $M_{\rm BH,\min}\sim$4.3 \Msun, significantly
above the maximum NS mass, further supporting the existence of a gap.
Theoretical interpretation of the observed mass distribution was given in the context of different supernova mechanisms \citep{BWF12}:
the gap appeared in the simulations with rapidly developing explosions
(launched within $\sim$0.2~s after the core bounce),
while the delayed explosions (developing on timescales $\sim$0.5--1~s) result in continuous mass distribution.
The obtained range of masses for \source\ put the binary right into the 2--5 \Msun\ gap, suggesting that the object might be formed in
the delayed explosion scenario.
Alternatively, the primary mass, initially below 2 \Msun, could have been enhanced as a result of the accretion processes on the Hubble
timescale, to reach the maximum possible mass of a stable NS and then collapsed into a BH.
Such effects of binary evolution were considered in \citet{BWF12}, who showed that this results in the small additional number of
objects with the masses in the range 2--3 \Msun.

On the other hand, investigations of \citet{KBF12} suggest that the gap is an artefact of systematic uncertainties in mass
measurements.
They conclude that the BH masses in two objects, GRO~J0422+32 and 4U~1543--47, plausibly lie within the mass gap.
Other `outliers' include the binary 4U~1700--37, $M=2.44\pm0.27$\Msun\ \citep{Clark02}, for a long time being
classified as a NS \citep{ROK99}, but again returning to the originally proposed \citep[e.g.][]{BWW96} BH
class by the discovery of the very low-frequency QPOs \citep{Dolan11}.
The compact object in the classical BH candidate Cyg~X-3 was recently estimated to
have $M=2.4^{+2.1}_{-1.1}$ \Msun\ \citep{AAZ13},  again right in the middle of the mass gap.
Our estimates for \source\ further support the existence of compact objects with masses in the range 2--5 \Msun.
A re-analysis of the full set of data is probably required to conclude whether the  mass gap  still exists.

Masses of BHs used to be comfortably above the limit of 3 \Msun.
However, a growing amount of measurements suggest that there exist a population of
compact objects very close to this limit.
They can be formed either in the processes of the delayed supernova explosion \citep[such as in simulations of][]{BWF12} or by the
accretion-induced collapse of the NS (initially formed in the supernova explosion) during the common-envelope phase or
by accretion from the companion on the Hubble time.
It becomes evident that the firm classification based on the compact object mass estimate alone is becoming less and less reliable, and
other diagnostic method should be agreed on to serve a criterion to distinguish between BHs and NSs.
Such a technique will likely be based on the difference of observational appearance of objects with and without solid surface.
The most promising methods are based on the X-ray colour evolution as a function of flux \citep{DG03} and on the
properties of the broad-band noise \citep{SR00}.

\begin{figure}
\includegraphics[angle=0,width=8.0cm]{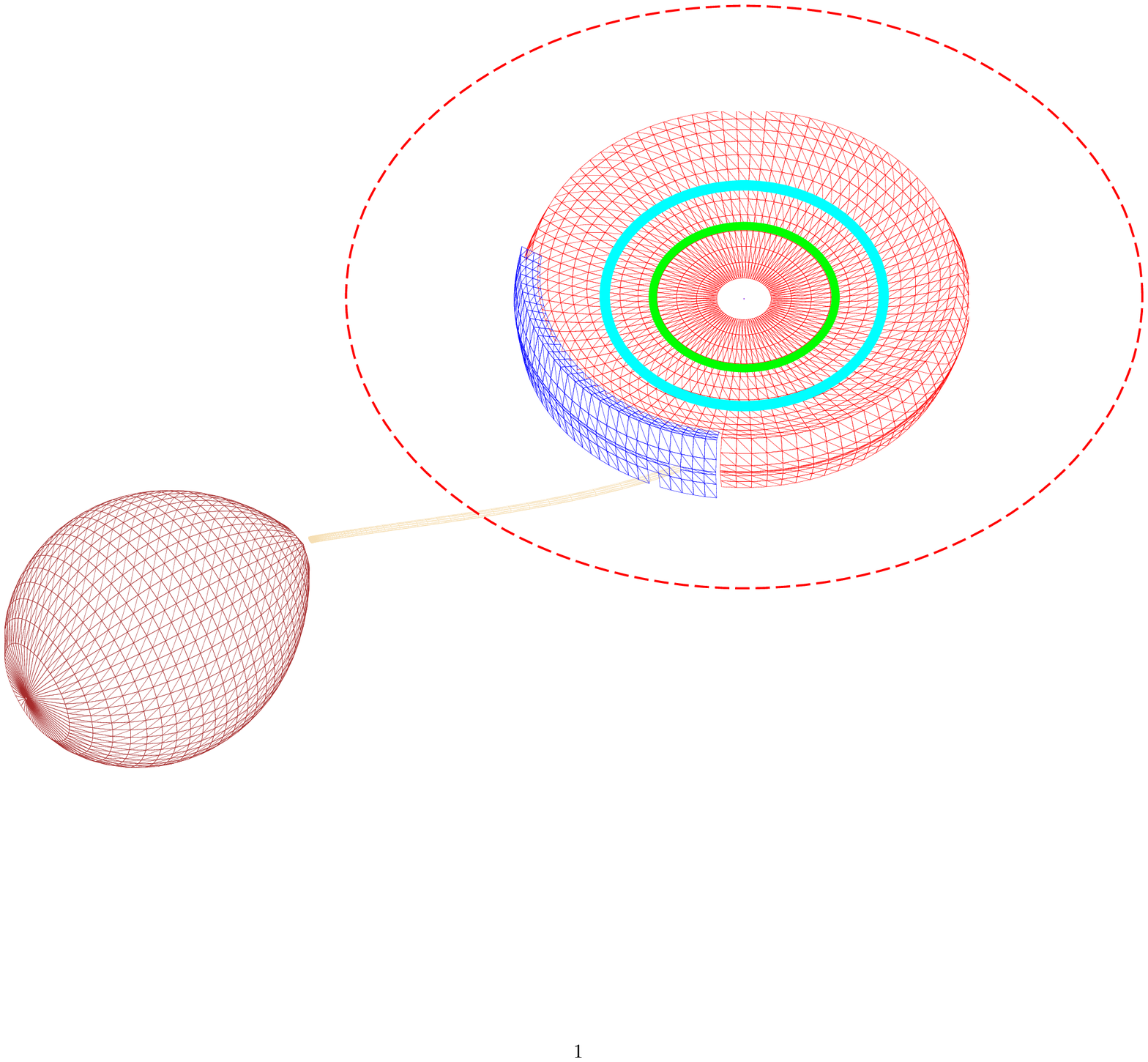}
%\vspace{5 mm}
\caption{A schematic representation of the suggested geometry for \source\ plotted using the system
parameters: $M_1$=3\Msun, $M_2$=0.2\Msun, $i$=40\degr. The red thick dashed
circle shows the largest radius of the accretion disc determined by tidal limitations. We suggest
that it was the accretion disc outer radius during the observations of \citet{Torres05b}. At the time
of our optical observations the disc (shown as a red disc) has shrunk by about 45 per cent compared
to Torres et al.'s observations. The cyan and green rings represent the disc outer radii in which
emission lines with higher excitation energies than \Halpha\ were originated at the time of
observations (cyan: \HeII\ $\lambda$4686 and \CIV, green: \HeII\ $\lambda$1640). The blue segment
located at the outer rim of the accretion disc in front of the secondary shows the bright area
which position is assigned according to the relative phasing of the photometric and radial velocity
modulations.
}
\label{Fig:Model}
\end{figure}

\subsection{Orbital period}

Both our photometric and spectroscopic data suggest the orbital period of \source\ 2.85~h. This
puts the system right into the so-called period gap between about 2.15 and 3.18~h
\citep[according to results of][]{Knigge06}. The gap marks the dearth of active CVs in this
period range, which appear to have smaller radii compared to the Roche lobe size. The traditional
explanation of the period gap involves a thermal bloating of the secondary at periods above
$\sim$3~h due to enhanced mass transfer rates (due to magnetic braking mechanism) over those
driven by the gravitational radiation losses alone \citep{HNR01,KBP11}. The magnetic braking
abruptly stops at this upper limit when the secondary becomes fully convective \citep{SR83},
leading to a reduction of the mass-loss rate and, as a result, its contraction. The binary
continues to evolve towards shorter periods as detached system, and eventually the Roche lobe
becomes small enough to resume the mass transfer. The fact that \source\ displays an outburst
whilst having the period in the middle of the gap supports our suggestion that the secondary is
a slightly evolved star, whose radius is larger than that of a main-sequence star of the same mass.
The same conclusion was reached for the LMXB MAXI~J1659--152 \citep[2.4~h,][]{Kuulkers13} and
can also be applied to SWIFT~J1357.2--0933 \citep[2.8~h,][]{corral-santana}.

\section{Summary}

Despite of the long history of study, system parameters of \source\, apart from the orbital period
were not investigated. The system exhibits X-ray properties typical for the BH binary and has
always been considered in the context of the BH primary. Its mass, however, was not dynamically
measured, and the assumed values in some cases reached 12 \Msun\ \citep{Zurita08,Froning}. For the
first time, we performed the analysis of spectroscopic data to determine the system parameters. Our
main result is that the primary in \source\ is not at all so massive -- its mass in excess of
5 \Msun\ is highly improbable. This result not only
affects the modelling of spectral properties, but also supports the possibility of existence of
compact objects in the range 2--5 \Msun, the so-called mass gap. The later conclusion greatly limits
the formation scenarios. For instance, in the rapid supernova explosion mechanisms, it is not
possible to produce the compact objects with masses 3--5 \Msun, even accounting for the binary
evolution effects. We suggest that the primary in \source\ can either be produced in the delayed
explosion or be the result of the accretion-induced collapse of a NS.

We also performed time series analysis of our photometric and spectroscopic data and confirm that
\source\ is one of the shortest period X-ray binaries, as initially proposed by \citet{Zurita08}.
Our data, however, are better described by the orbital period of 2.85~h, in contrast to $\sim$3.24~h
found by them. This finding puts the system into the well-known period gap, in which the
main-sequence companions do not fill their Roche lobes. The fact that \source\ is still in outburst
suggests that the mass transfer proceeds, therefore, the companion has a radius larger than that of
a main sequence star of the same mass. This is naturally explained if the secondary is somewhat
nuclearly evolved.

\section*{Acknowledgments}

This work was supported by the Finnish  Doctoral Program in Astronomy and Space Physics
(AV) and the Academy of Finland grant 268740 (JP). SZ acknowledges PAPIIT grants IN-100614
and CONACyT grants 151858 and CAR 208512 for resources provided toward this research.
JJEK acknowledges partial financial support from the Emil Aaltonen Foundation and
the V\"{a}is\"{a}l\"{a} Foundation.
Our research was partially based on UV observations by NASA missions \emph{HST} and \emph{Swift}
which we acknowledge. We thank Neil Gehrels for approving the Target of Opportunity
observation with \emph{Swift} and the \emph{Swift} team for executing the observation. This research
has made use of data obtained through the High Energy Astrophysics Science Archive
Research Center Online Service, provided by the NASA/Goddard Space Flight Center. We
would like to acknowledge the anonymous referee whose comments have significantly
improved this paper.

\appendix

\section{Brief description of the light-curve simulations}
\label{Appendix}

The detailed fitting of the light curve of \source\ requires to make certain assumptions on the
accretion disc structure which we do not know. In order to reproduce the observed light curve of
\source\ in terms of shape and amplitude, we adopted the modelling technique from \citet{ADModel2}
and \citet{ADModel}. A simple geometrical model of \source, presented in Fig.~\ref{Fig:Model} is
comprised of a concave accretion disc, a secondary red dwarf star, a stream from the inner Lagrangian
point, and a bright area located at the outer rim of the accretion disc in front of the secondary.

We assume that the steady-state, optically thick, viscous accretion disc radiates as a blackbody at
the local temperature which radial distribution across the disc is given by
$T(r) = T_{\rm out}\,(r/r_{\rm out})^{-3/4}$,
where $T_{\rm out}$ is the temperature at the outer edge of the disc. The bright area is characterized
by the azimuthal extension $\theta$ and the temperature $T_{\rm b}$. The vertical extension of the bright
area is equal to the thickness of the concave accretion disc at the outer edge and characterized by the
opening angle $z$ as seen from the BH. The donor star fills its Roche lobe. Emission from the accretion
stream is not taken into account. The surface of each component of the system is divided
in a series of triangles as shown in Fig.~\ref{Fig:Model}, each triangle emits as a blackbody with
corresponding temperature. The total flux from the binary is obtained by integrating the emission
from all the elements lying in a sight of view and then folded with the response of the $V$ filter.

The fixed parameters for the model are set as follows:
\begin{enumerate}
  \item The mass of the primary is 3\Msun;
  \item the mass of the donor star is 0.2\Msun;
  \item the temperature of the donor star is 3000K;
  \item the outer and inner radii of the disc are imposed by the parameters $V_{\rm out}=800$ \kms\
        and $r_{\rm in}/r_{\rm out}=0.02$ estimated for the \Halpha\ emission line (see
        Section~\ref{Sec:modelling} and Table~\ref{LineParamTab});
  \item the temperature at the outer edge of the disc is $T_{\rm out}$=8\,000K.
\end{enumerate}

\begin{figure}
  \includegraphics[height=6.5cm]{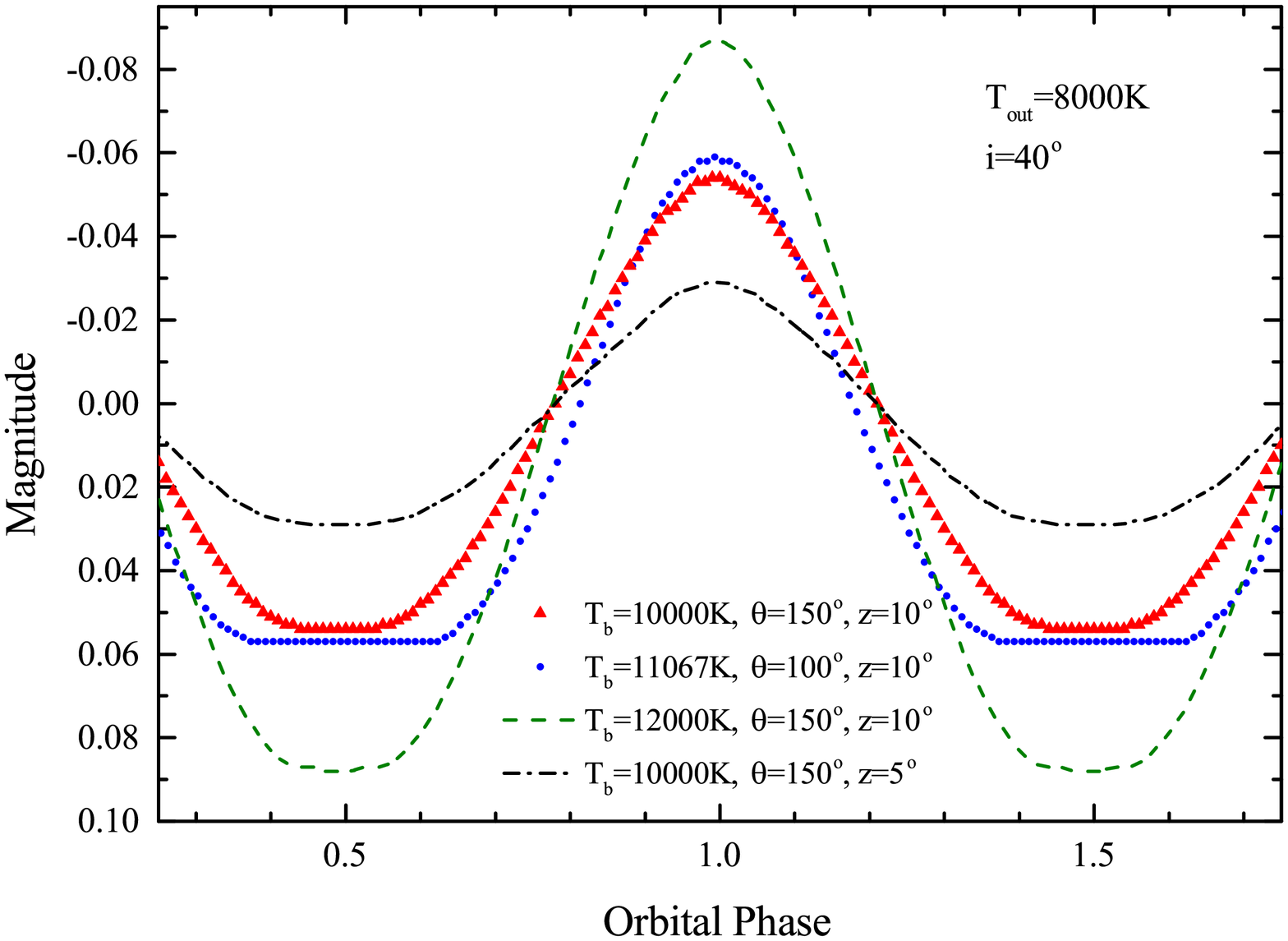}
\caption{Light curves calculated for different bright area parameters.
}
\label{Fig:LCmodels1}
\end{figure}

\begin{figure}
  \includegraphics[height=6.5cm]{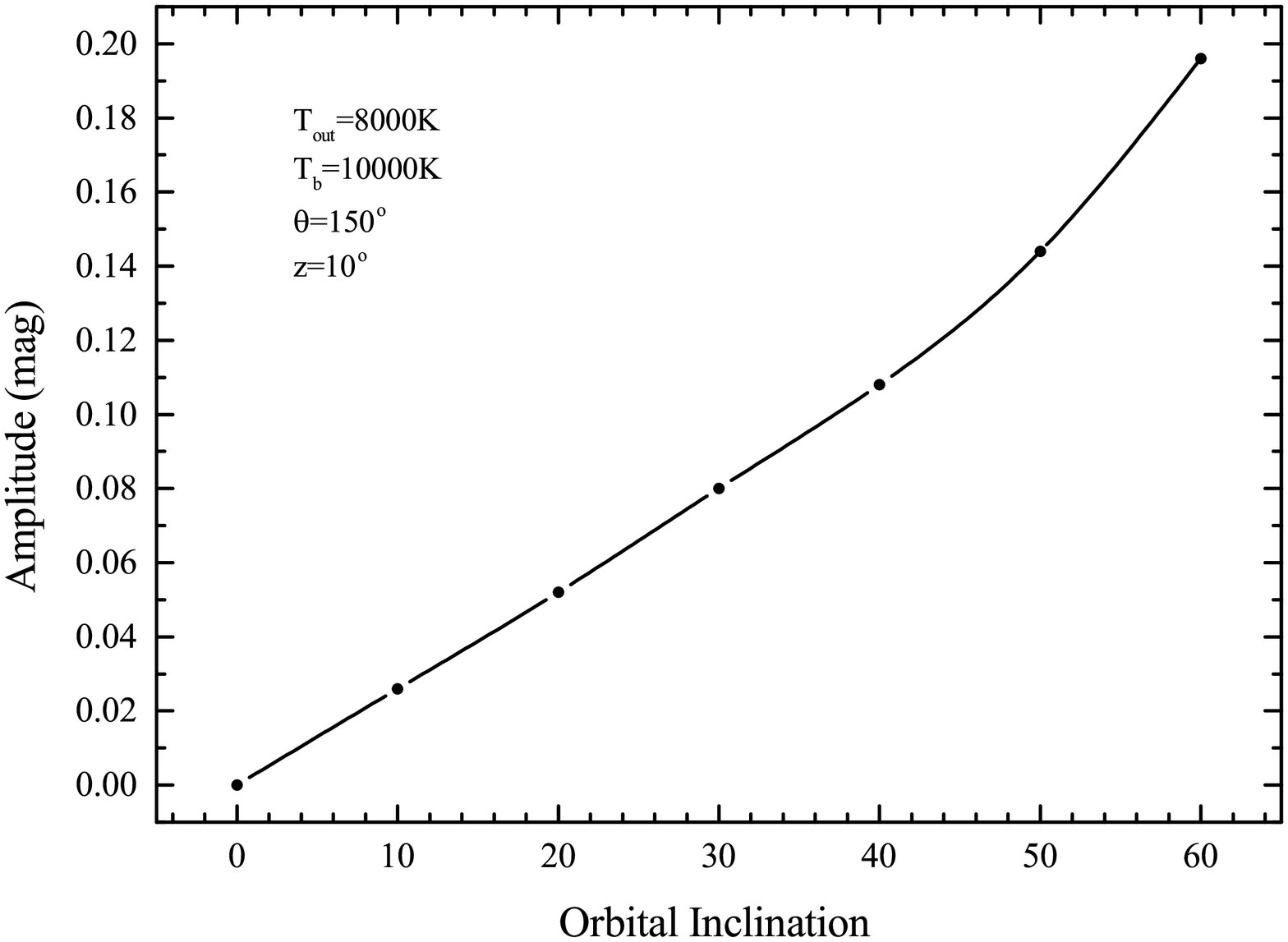}
\caption{The dependence of the full amplitude of orbital modulation on the orbital
inclination.}
\label{Fig:LCmodels2}
\end{figure}

We calculated a variety of models using different bright area parameters.
We found that the shape of the light curve mostly depends on the azimuthal extension $\theta$
whereas the amplitude of variability relies on the orbital inclination $i$, the opening
angle of the disc $z$, and the ratio of temperatures of the bright area and the disc
$T_{\rm b}/T_{\rm out}$ (Fig.~\ref{Fig:LCmodels1}).
The amplitude of light curve decreases quickly with decreasing $i$ (Fig.~\ref{Fig:LCmodels2}).
Our simulations show that for the orbital inclination of 30\degr\ the bright
area with $T_{\rm b}$=10\,000K, $\theta$=150\degr\ and $z$=10\degr\ and the accretion disc
with the above parameters creates the modulation with the total amplitude of only 0.08 mag,
less than the observed amplitude. Here we claim that these parameters of the bright area
are \textit{unrealistic}. They correspond to the blackbody luminosity of the bright area
which is twice as large as the total luminosity of the visible half of the accretion disc
rim that is not observed in LMXBs or CVs in outbursts. Furthermore, the decrease of the
geometrical parameters of the bright area will require to increase this discrepancy further
in order to keep the same amplitude of the orbital modulation.

%-----------------------------Figure Start------------------------------
%-----------------------------Figure End--------------------------------

%-----------------------------Figure Start------------------------------
%-----------------------------Figure End--------------------------------

Thus, we conclude that with the use of more or less realistic parameters for the bright area
it is nearly impossible to obtain the required orbital variability with the amplitude of at
least 0.10 mag for the orbital inclination $i\lesssim$40\degr. Only extraordinarily bright
and extended area can produce such variability.

\label{lastpage}

\begin{thebibliography}{84}
\expandafter\ifx\csname natexlab\endcsname\relax\def\natexlab#1{#1}\fi


\bibitem[{{Bailyn} {et~al}\mbox{.}(1998){Bailyn}, {Jain}, {Coppi}, \&
  {Orosz}}]{Bailyn98}
{Bailyn} C.~D., {Jain} R.~K., {Coppi} P., {Orosz} J.~A., 1998, \apj, 499, 367

\bibitem[\protect\citeauthoryear{Barman, Hauschildt, \& Allard}{2004}]{Barman}
  Barman T.~S., Hauschildt P.~H., Allard F., 2004, ApJ, 614, 338

\bibitem[{{Barret} {et~al}\mbox{.}(2000){Barret}, {Olive}, {Boirin}, {Done},
  {Skinner}, \& {Grindlay}}]{Barret00}
{Barret} D., {Olive} J.~F., {Boirin} L., {Done} C., {Skinner} G.~K., {Grindlay}
  J.~E., 2000, \apj, 533, 329

\bibitem[{{Belczynski} {et~al}\mbox{.}(2012){Belczynski}, {Wiktorowicz},
  {Fryer}, {Holz}, \& {Kalogera}}]{BWF12}
{Belczynski} K., {Wiktorowicz} G., {Fryer} C.~L., {Holz} D.~E., {Kalogera} V.,
  2012, \apj, 757, 91

\bibitem[{{Belloni} {et~al}\mbox{.}(2011){Belloni}, {Motta}, \&
  {Mu{\~n}oz-Darias}}]{Belloni11}
{Belloni} T.~M., {Motta} S.~E., {Mu{\~n}oz-Darias} T., 2011, Bull. Astr. Soc. India, 39, 409

\bibitem[{{Berry} \& {Burnell}(2005)}]{AIP4Win}
{Berry} R., {Burnell} J., 2005, {The handbook of astronomical image processing,
  2nd ed.} Willmann-Bell, Richmond, VA

\bibitem[{{Borisov} \& {Neustroev}(1998)}]{BorisovNeustroev}
{Borisov} N.~V., {Neustroev} V.~V., 1998, Bull. Spec. Astrophys. Obs., 44, 110 (astro-ph/9806159)

\bibitem[{{Brown} {et~al}\mbox{.}(1996){Brown}, {Weingartner}, \&
  {Wijers}}]{BWW96}
{Brown} G.~E., {Weingartner} J.~C., {Wijers} R.~A.~M.~J., 1996, \apj, 463, 297

\bibitem[{{Butler} {et~al}\mbox{.}(2012){Butler}, {Klein}, {Fox}, {Lotkin},
  {Bloom}, {Prochaska}, {Ramirez-Ruiz}, {de Diego}, {Georgiev}, {Gonz{\'a}lez},
  {Lee}, {Richer}, {Rom{\'a}n}, {Watson}, {Gehrels}, {Kutyrev}, {Bernstein},
  {Alvarez}, {Cese{\~n}a}, {Clark}, {Colorado}, {C{\'o}rdova}, {Farah},
  {Garc{\'{\i}}a}, {Guisa}, {Herrera}, {Lazo}, {L{\'o}pez}, {Luna},
  {Mart{\'{\i}}nez}, {Murillo}, {Murillo}, {N{\'u}{\~n}ez}, {Pedrayes},
  {Quir{\'o}s}, {Ochoa}, {Sierra}, {Moseley}, {Rapchun}, {Robinson}, {Samuel},
  \& {Sparr}}]{Ratir1}
{Butler} N. {et~al.}, 2012, SPIE, 844610

\bibitem[{{Cadolle Bel} {et~al}\mbox{.}(2007){Cadolle Bel}, {Rib{\'o}},
  {Rodriguez}, {Chaty}, {Corbel}, {Goldwurm}, {Frontera}, {Farinelli},
  {D'Avanzo}, {Tarana}, {Ubertini}, {Laurent}, {Goldoni}, \&
  {Mirabel}}]{Cadolle07}
{Cadolle Bel} M. {et~al.}, 2007, \apj, 659, 549

\bibitem[{{Calvelo} {et~al}\mbox{.}(2009){Calvelo}, {Vrtilek}, {Steeghs},
  {Torres}, {Neilsen}, {Filippenko}, \& {Gonz{\'a}lez
  Hern{\'a}ndez}}]{XTEJ1118}
{Calvelo} D.~E., {Vrtilek} S.~D., {Steeghs} D., {Torres} M.~A.~P., {Neilsen}
  J., {Filippenko} A.~V., {Gonz{\'a}lez Hern{\'a}ndez} J.~I., 2009, \mnras,
  399, 539

\bibitem[\protect\citeauthoryear{Casares \& Jonker}{2014}]{CasaresJonker} 
    Casares J., Jonker P.~G., 2014, SSRv, 183, 223


\bibitem[{{Casares} {et~al}\mbox{.}(2003){Casares}, {Steeghs}, {Hynes},
  {Charles}, \& {O'Brien}}]{CSH03}
{Casares} J., {Steeghs} D., {Hynes} R.~I., {Charles} P.~A., {O'Brien} K., 2003,
  \apj, 590, 1041

\bibitem[{{Chiang} {et~al}\mbox{.}(2010){Chiang}, {Done}, {Still}, \&
  {Godet}}]{Chiang10}
{Chiang} C.~Y., {Done} C., {Still} M., {Godet} O., 2010, \mnras, 403, 1102

\bibitem[{{Clark} {et~al}\mbox{.}(2002){Clark}, {Goodwin}, {Crowther}, {Kaper},
  {Fairbairn}, {Langer}, \& {Brocksopp}}]{Clark02}
{Clark} J.~S., {Goodwin} S.~P., {Crowther} P.~A., {Kaper} L., {Fairbairn} M.,
  {Langer} N., {Brocksopp} C., 2002, \aap, 392, 909

\bibitem[{{Corral-Santana} {et~al}\mbox{.}(2013){Corral-Santana}, {Casares},
  {Mu{\~n}oz-Darias}, {Rodr{\'{\i}}guez-Gil}, {Shahbaz}, {Torres}, {Zurita}, \&
  {Tyndall}}]{corral-santana}
{Corral-Santana} J.~M., {Casares} J., {Mu{\~n}oz-Darias} T.,
  {Rodr{\'{\i}}guez-Gil} P., {Shahbaz} T., {Torres} M.~A.~P., {Zurita} C.,
  {Tyndall} A.~A., 2013, Sci, 339, 1048

%\bibitem[\protect\citeauthoryear{Cowley et al.}{2002}]{CowleyJet} Cowley A.~P.,
%  Schmidtke P.~C., Hutchings J.~B., Crampton D., 2002, AJ, 123, 1741


\bibitem[{{Dolan}(2011)}]{Dolan11}
{Dolan} J.~F., 2011, arXiv:1107.1537

\bibitem[{{Done} \& {Gierli{\'n}ski}(2003)}]{DG03}
{Done} C., {Gierli{\'n}ski} M., 2003, \mnras, 342, 1041

\bibitem[\protect\citeauthoryear{Dubus et al.}{1999}]{Dubus} Dubus G., Lasota J.-P.,
  Hameury J.-M., Charles P., 1999, MNRAS, 303, 139

\bibitem[{{Durant} {et~al}\mbox{.}(2009){Durant}, {Gandhi}, {Shahbaz},
  {Peralta}, \& {Dhillon}}]{Durant09}
{Durant} M., {Gandhi} P., {Shahbaz} T., {Peralta} H.~H., {Dhillon} V.~S., 2009,
  \mnras, 392, 309

\bibitem[{{Eggleton}(1983)}]{Eggleton83}
{Eggleton} P.~P., 1983, \apj, 268, 368

\bibitem[{{Farr} {et~al}\mbox{.}(2011){Farr}, {Sravan}, {Cantrell},
  {Kreidberg}, {Bailyn}, {Mandel}, \& {Kalogera}}]{BH-Farr}
{Farr} W.~M., {Sravan} N., {Cantrell} A., {Kreidberg} L., {Bailyn} C.~D.,
  {Mandel} I., {Kalogera} V., 2011, \apj, 741, 103

\bibitem[{{Froning} {et~al}\mbox{.}(2014){Froning}, {Maccarone}, {France},
  {Winter}, {Robinson}, {Hynes}, \& {Lewis}}]{Froning}
{Froning} C.~S., {Maccarone} T.~J., {France} K., {Winter} L., {Robinson} E.~L.,
  {Hynes} R.~I., {Lewis} F., 2014, \apj, 780, 48

\bibitem[{{Gehrels} {et~al}\mbox{.}(2004){Gehrels}, {Chincarini}, {Giommi},
  {Mason}, {Nousek}, {Wells}, {White}, {Barthelmy}, {Burrows}, {Cominsky},
  {Hurley}, {Marshall}, {M{\'e}sz{\'a}ros}, {Roming}, {Angelini}, {Barbier},
  {Belloni}, {Campana}, {Caraveo}, {Chester}, {Citterio}, {Cline}, {Cropper},
  {Cummings}, {Dean}, {Feigelson}, {Fenimore}, {Frail}, {Fruchter}, {Garmire},
  {Gendreau}, {Ghisellini}, {Greiner}, {Hill}, {Hunsberger}, {Krimm},
  {Kulkarni}, {Kumar}, {Lebrun}, {Lloyd-Ronning}, {Markwardt}, {Mattson},
  {Mushotzky}, {Norris}, {Osborne}, {Paczynski}, {Palmer}, {Park}, {Parsons},
  {Paul}, {Rees}, {Reynolds}, {Rhoads}, {Sasseen}, {Schaefer}, {Short},
  {Smale}, {Smith}, {Stella}, {Tagliaferri}, {Takahashi}, {Tashiro},
  {Townsley}, {Tueller}, {Turner}, {Vietri}, {Voges}, {Ward}, {Willingale},
  {Zerbi}, \& {Zhang}}]{Swift}
{Gehrels} N. {et~al.}, 2004, \apj, 611, 1005

\bibitem[{{Gierli{\'n}ski} \& {Poutanen}(2005)}]{GP05}
{Gierli{\'n}ski} M., {Poutanen} J., 2005, \mnras, 359, 1261

\bibitem[{{Gilfanov}(2010)}]{Gilfanov10}
{Gilfanov} M., 2010, in Lecture Notes in Physics, Vol. 794, The Jet Paradigm,
  {T.~Belloni}, ed., Springer-Verlag, Berlin, pp.~17--52

\bibitem[{{Horne} \& {Marsh}(1986)}]{HorneMarsh}
{Horne} K., {Marsh} T.~R., 1986, \mnras, 218, 761

\bibitem[{{Horne} \& {Saar}(1991)}]{HorneSaar}
{Horne} K., {Saar} S.~H., 1991, \apjl, 374, L55

\bibitem[{{Howell} {et~al}\mbox{.}(2001){Howell}, {Nelson}, \&
  {Rappaport}}]{HNR01}
{Howell} S.~B., {Nelson} L.~A., {Rappaport} S., 2001, \apj, 550, 897

\bibitem[{{Hynes} {et~al}\mbox{.}(2003){Hynes}, {Steeghs}, {Casares},
  {Charles}, \& {O'Brien}}]{HSC03}
{Hynes} R.~I., {Steeghs} D., {Casares} J., {Charles} P.~A., {O'Brien} K., 2003,
  \apjl, 583, L95

\bibitem[{{Ibragimov} \& {Poutanen}(2009)}]{IP09}
{Ibragimov} A., {Poutanen} J., 2009, \mnras, 400, 492

\bibitem[{{Johnston} {et~al}\mbox{.}(1989){Johnston}, {Kulkarni}, \&
  {Oke}}]{JKO89}
{Johnston} H.~M., {Kulkarni} S.~R., {Oke} J.~B., 1989, \apj, 345, 492

\bibitem[{{Knigge}(2006)}]{Knigge06}
{Knigge} C., 2006, \mnras, 373, 484

\bibitem[{{Knigge}(2012)}]{Knigge_rev}
{Knigge} C., 2012, \memsai, 83, 549

\bibitem[{{Knigge} {et~al}\mbox{.}(2011){Knigge}, {Baraffe}, \&
  {Patterson}}]{KBP11}
{Knigge} C., {Baraffe} I., {Patterson} J., 2011, \apjs, 194, 28

\bibitem[{{Kolb} {et~al}\mbox{.}(2001){Kolb}, {King}, \& {Baraffe}}]{KKB01}
{Kolb} U., {King} A.~R., {Baraffe} I., 2001, \mnras, 321, 544

\bibitem[{{Kreidberg} {et~al}\mbox{.}(2012){Kreidberg}, {Bailyn}, {Farr}, \&
  {Kalogera}}]{KBF12}
{Kreidberg} L., {Bailyn} C.~D., {Farr} W.~M., {Kalogera} V., 2012, \apj, 757,
  36

\bibitem[{{Kuulkers} {et~al}\mbox{.}(2013){Kuulkers}, {Kouveliotou}, {Belloni},
  {Cadolle Bel}, {Chenevez}, {D{\'{\i}}az Trigo}, {Homan}, {Ibarra}, {Kennea},
  {Mu{\~n}oz-Darias}, {Ness}, {Parmar}, {Pollock}, {van den Heuvel}, \& {van
  der Horst}}]{Kuulkers13}
{Kuulkers} E. {et~al.}, 2013, \aap, 552, A32

\bibitem[{{La Dous}(1989)}]{LaDous}
{La Dous} C., 1989, \aap, 211, 131

\bibitem[{{Lattimer}(2012)}]{Lat12}
{Lattimer} J.~M., 2012, ARNPS, 62, 485

\bibitem[{{Lin} {et~al}\mbox{.}(2007){Lin}, {Remillard}, \& {Homan}}]{Lin07}
{Lin} D., {Remillard} R.~A., {Homan} J., 2007, \apj, 667, 1073

\bibitem[{{Lin} {et~al}\mbox{.}(2010){Lin}, {Remillard}, \& {Homan}}]{Lin10}
{Lin} D., {Remillard} R.~A., {Homan} J., 2010, \apj, 719, 1350

\bibitem[{{Lin} {et~al}\mbox{.}(2011){Lin}, {Rappaport}, {Podsiadlowski},
  {Nelson}, {Paxton}, \& {Todorov}}]{LRP11}
{Lin} J., {Rappaport} S., {Podsiadlowski} P., {Nelson} L., {Paxton} B.,
  {Todorov} P., 2011, \apj, 732, 70

\bibitem[\protect\citeauthoryear{Marsh}{2001}]{marsh2001} Marsh T.~R., 2001,
   in Astrotomography, Indirect Imaging Methods in Observational Astronomy,
   ed. H.~M.~J.~Boffin, D.~Steeghs, and J.~Cuypers, Lect. Notes Phys., 573, 1

\bibitem[{{Marsh} \& {Horne}(1988)}]{Marsh-Horne-88}
  {Marsh} T.~R., {Horne} K., 1988, \mnras, 235, 269

\bibitem[\protect\citeauthoryear{Marsh \& Horne}{1990}]{Marsh-Horne-90} Marsh T.~R.,
  Horne K., 1990, ApJ, 349, 593

\bibitem[{{Massa} {et~al}\mbox{.}(2010){Massa}, {Aloisi}, {Keyes}, {Bohlin}, \&
  {Froning}}]{COSreport}
{Massa} D., {Aloisi} A., {Keyes} C., {Bohlin} R., {Froning} C., 2010,
Instrument Science Report COS 2010-01(v1). Space Telescope Science Institute

\bibitem[{{Morgan} {et~al}\mbox{.}(2005){Morgan}, {Swank}, {Markwardt}, \&
  {Gehrels}}]{Morgan05}
{Morgan} E., {Swank} J., {Markwardt} C., {Gehrels} N., 2005, ATel, 550, 1

\bibitem[{{Mu{\~n}oz-Darias} {et~al}\mbox{.}(2008){Mu{\~n}oz-Darias},
  {Casares}, \& {Mart{\'{\i}}nez-Pais}}]{munoz08}
{Mu{\~n}oz-Darias} T., {Casares} J., {Mart{\'{\i}}nez-Pais} I.~G., 2008,
  \mnras, 385, 2205

\bibitem[{{Neustroev}(1998)}]{NeustroevWZ}
{Neustroev} V.~V., 1998, Astr. Rep., 42, 748

\bibitem[{{Neustroev} {et~al}\mbox{.}(2002){Neustroev}, {Borisov}, {Barwig},
  {Bobinger}, {Mantel}, {{\v S}imi{\'c}}, \& {Wolf}}]{NeustroevIP}
{Neustroev} V.~V., {Borisov} N.~V., {Barwig} H., {Bobinger} A., {Mantel} K.~H.,
  {{\v S}imi{\'c}} D., {Wolf} S., 2002, \aap, 393, 239

\bibitem[{{Neustroev} \& {Zharikov}(2008)}]{BF_Eri}
{Neustroev} V.~V., {Zharikov} S., 2008, \mnras, 386, 1366

\bibitem[{{Oke}(1990)}]{Oke}
{Oke} J.~B., 1990, \aj, 99, 1621

\bibitem[{{Orosz} {et~al}\mbox{.}(1994){Orosz}, {Bailyn}, {Remillard},
  {McClintock}, \& {Foltz}}]{Orosz94}
{Orosz} J.~A., {Bailyn} C.~D., {Remillard} R.~A., {McClintock} J.~E., {Foltz}
  C.~B., 1994, \apj, 436, 848

\bibitem[{{Orosz} {et~al}\mbox{.}(2002){Orosz}, {Groot}, {van der Klis},
  {McClintock}, {Garcia}, {Zhao}, {Jain}, {Bailyn}, \& {Remillard}}]{Orosz02}
{Orosz} J.~A. {et~al.}, 2002, \apj, 568, 845

\bibitem[{{{\"O}zel} {et~al}\mbox{.}(2010){{\"O}zel}, {Psaltis}, {Narayan}, \&
  {McClintock}}]{OzelBH}
{{\"O}zel} F., {Psaltis} D., {Narayan} R., {McClintock} J.~E., 2010, \apj, 725, 1918

\bibitem[{{Palmer} {et~al}\mbox{.}(2005){Palmer}, {Barthelmey}, {Cummings},
  {Gehrels}, {Krimm}, {Markwardt}, {Sakamoto}, \& {Tueller}}]{Palmer05}
{Palmer} D.~M., {Barthelmey} S.~D., {Cummings} J.~R., {Gehrels} N., {Krimm}
  H.~A., {Markwardt} C.~B., {Sakamoto} T., {Tueller} J., 2005, ATel, 546, 1

\bibitem[{{Patterson} {et~al}\mbox{.}(2005){Patterson}, {Kemp}, {Harvey},
  {Fried}, {Rea}, {Monard}, {Cook}, {Skillman}, {Vanmunster}, {Bolt},
  {Armstrong}, {McCormick}, {Krajci}, {Jensen}, {Gunn}, {Butterworth}, {Foote},
  {Bos}, {Masi}, \& {Warhurst}}]{Patterson05}
{Patterson} J. {et~al.}, 2005, \pasp, 117, 1204

\bibitem[{{Podsiadlowski} {et~al}\mbox{.}(2002){Podsiadlowski}, {Rappaport}, \&
  {Pfahl}}]{PRP02}
{Podsiadlowski} P., {Rappaport} S., {Pfahl} E.~D., 2002, \apj, 565, 1107

\bibitem[{{Poole} {et~al}\mbox{.}(2008){Poole}, {Breeveld}, {Page}, {Landsman},
  {Holland}, {Roming}, {Kuin}, {Brown}, {Gronwall}, {Hunsberger}, {Koch},
  {Mason}, {Schady}, {vanden Berk}, {Blustin}, {Boyd}, {Broos}, {Carter},
  {Chester}, {Cucchiara}, {Hancock}, {Huckle}, {Immler}, {Ivanushkina},
  {Kennedy}, {Marshall}, {Morgan}, {Pandey}, {de Pasquale}, {Smith}, \&
  {Still}}]{UVOT}
{Poole} T.~S. {et~al.}, 2008, \mnras, 383, 627

\bibitem[{{Poutanen} \& {Gierli{\'n}ski}(2003)}]{PG03}
{Poutanen} J., {Gierli{\'n}ski} M., 2003, \mnras, 343, 1301

\bibitem[{{Puebla} {et~al}\mbox{.}(2011){Puebla}, {Diaz}, {Hillier}, \&
  {Hubeny}}]{Puebla2011}
{Puebla} R.~E., {Diaz} M.~P., {Hillier} D.~J., {Hubeny} I., 2011, \apj, 736, 17

\bibitem[{{Puebla} {et~al}\mbox{.}(2007){Puebla}, {Diaz}, \&
  {Hubeny}}]{Puebla2007}
{Puebla} R.~E., {Diaz} M.~P., {Hubeny} I., 2007, \aj, 134, 1923

\bibitem[{{Remillard} \& {McClintock}(2006)}]{RM06}
{Remillard} R.~A., {McClintock} J.~E., 2006, \araa, 44, 49

\bibitem[{{Reynolds} {et~al}\mbox{.}(1999){Reynolds}, {Owens}, {Kaper},
  {Parmar}, \& {Segreto}}]{ROK99}
{Reynolds} A.~P., {Owens} A., {Kaper} L., {Parmar} A.~N., {Segreto} A., 1999,
  \aap, 349, 873

\bibitem[\protect\citeauthoryear{Ritter}{2008}]{Ritter} Ritter H., 2008, NewAR, 51, 869

\bibitem[{{Schneider} \& {Young}(1980)}]{sch:young}
{Schneider} D.~P., {Young} P., 1980, \apj, 238, 946

\bibitem[{{Shafter}(1983)}]{Shafter}
{Shafter} A.~W., 1983, \apj, 267, 222

\bibitem[{{Smak}(1981)}]{Smak1981}
{Smak} J., 1981, \actaa, 31, 395

\bibitem[\protect\citeauthoryear{Smith \& Dhillon}{1998}]{Smith:Dhillon} Smith D.~A.,
  Dhillon V.~S., 1998, MNRAS, 301, 767

\bibitem[{{Soleri} {et~al}\mbox{.}(2013){Soleri}, {Mu{\~n}oz-Darias}, {Motta},
  {Belloni}, {Casella}, {M{\'e}ndez}, {Altamirano}, {Linares}, {Wijnands},
  {Fender}, \& {van der Klis}}]{Soleri13}
{Soleri} P. {et~al.}, 2013, \mnras, 429, 1244

\bibitem[\protect\citeauthoryear{Soria, Wu, \& Johnston}{1999}]{Line6507}
  Soria R., Wu K., Johnston H.~M., 1999, MNRAS, 310, 71

\bibitem[{{Spruit}(1998)}]{Spruit}
{Spruit} H.~C., 1998, preprint (astro-ph/9806141)

\bibitem[{{Spruit} \& {Ritter}(1983)}]{SR83}
{Spruit} H.~C., {Ritter} H., 1983, \aap, 124, 267

\bibitem[{{Steeghs} \& {Casares}(2002)}]{SC02}
{Steeghs} D., {Casares} J., 2002, \apj, 568, 273

\bibitem[{{Sunyaev} \& {Revnivtsev}(2000)}]{SR00}
{Sunyaev} R., {Revnivtsev} M., 2000, \aap, 358, 617

\bibitem[{{Torres} {et~al}\mbox{.}(2005){Torres}, {Steeghs}, {Blake}, {Jonker},
  {Garcia}, {McClintock}, {Miller}, {Zhao}, {Calkins}, {Berlind}, {Falco},
  {Bloom}, {Callanan}, \& {Rodriguez-Gil}}]{Torres05b}
{Torres} M.~A.~P. {et~al.}, 2005, ATel, 566, 1

\bibitem[\protect\citeauthoryear{Tovmassian et al.}{2014}]{ADModel} Tovmassian G.,
  Stephania Hernandez M., Gonz{\'a}lez-Buitrago D., Zharikov S.,
  Garc{\'{\i}}a-D{\'{\i}}az M.~T., 2014, AJ, 147, 68

\bibitem[\protect\citeauthoryear{Wade \& Horne}{1988}]{WadeHorne} Wade R.~A., Horne K.,
  1988, ApJ, 324, 411

\bibitem[{{Warner}(1995)}]{Warner}
{Warner} B., 1995, Cataclysmic variable stars, Cambridge Astrophysics Series, No.  28.
Cambridge University Press, Cambridge

\bibitem[{{Watson} {et~al}\mbox{.}(2012){Watson}, {Richer}, {Bloom}, {Butler},
  {Cese{\~n}a}, {Clark}, {Colorado}, {C{\'o}rdova}, {Farah}, {Fox-Machado},
  {Fox}, {Garc{\'{\i}}a}, {Georgiev}, {Gonz{\'a}lez}, {Guisa}, {Guti{\'e}rrez},
  {Herrera}, {Klein}, {Kutyrev}, {Lazo}, {Lee}, {L{\'o}pez}, {Luna},
  {Mart{\'{\i}}nez}, {Murillo}, {Murillo}, {N{\'u}{\~n}ez}, {Prochaska},
  {Ochoa}, {Quir{\'o}s}, {Rapchun}, {Rom{\'a}n-Z{\'u}{\~n}iga}, \&
  {Valyavin}}]{Ratir2}
{Watson} A.~M. {et~al.}, 2012, SPIE, 84445L

\bibitem[{{Zdziarski} \& {Gierli{\'n}ski}(2004)}]{ZG04}
{Zdziarski} A.~A., {Gierli{\'n}ski} M., 2004, Prog. Theor. Phys. Suppl., 155, 99

\bibitem[{{Zdziarski} {et~al}\mbox{.}(2013){Zdziarski}, {Mikolajewska}, \&
  {Belczy{\'n}ski}}]{AAZ13}
{Zdziarski} A.~A., {Mikolajewska} J., {Belczy{\'n}ski} K., 2013, \mnras,
  429, L104

\bibitem[\protect\citeauthoryear{Zdziarski et al.}{1998}]{ZPM98} Zdziarski A.~A.,
  Poutanen J., Mikolajewska J., Gierlinski M., Ebisawa K., Johnson W.~N., 1998, MNRAS,
  301, 435

\bibitem[\protect\citeauthoryear{Zharikov et al.}{2013}]{ADModel2} Zharikov S., Tovmassian G.,
    Aviles A., Michel R., Gonzalez-Buitrago D., Garc{\'{\i}}a-D{\'{\i}}az M.~T., 2013, A\&A,
    549, A77

\bibitem[{{Zurita} {et~al}\mbox{.}(2008){Zurita}, {Durant}, {Torres},
  {Shahbaz}, {Casares}, \& {Steeghs}}]{Zurita08}
{Zurita} C., {Durant} M., {Torres} M.~A.~P., {Shahbaz} T., {Casares} J.,
  {Steeghs} D., 2008, \apj, 681, 1458

\end{thebibliography}
\end{document}